\documentclass{cup-hpl}
\usepackage[none]{hyphenat}
\usepackage{lipsum}

\begin{document}

\newtheorem{theorem}{Theorem}

\shorttitle{Design of an Active Thomson Parabola for the detection of ions accelerated by lasers}
\shortauthor{A. Huber et al.}

\title{Design of an Active Thomson Parabola for the detection of ions accelerated by laser: Numerical simulations and characterization of different solutions}

\author[1]{A. Huber\corresp{University of Bordeaux, CNRS, LP2I, UMR 5797, F-33170 Gradignan, France
                       \email{huber@lp2ib.in2p3.fr}}}
\author[1]  {M. Tarisien}                       
\author[1]  {F. Hannachi}
\author[2]  {M. Huault}
\author[1]  {J. Jouve}
\author[1]  {A. Maitrallain}
\author[2]  {P. Nicolaï}
\author[2]  {B. Zielbauer}
\author[2]  {D. Raffestin}

\address[1]{University of Bordeaux, CNRS, LP2I, UMR 5797, F-33170 Gradignan, France}
\address[2]{Université Bordeaux-CNRS-CEA, CELIA, UMR5107, 33405 Talence, France}
\address[3]{Plasma Physik/PHELIX, GSI Helmholtzzentrum für Schwerionenforschung GmbH, 64291 Darmstadt, Germany}

\begin{abstract}
This article presents the development of a modulable and active Thomson Parabola ion spectrometer designed to measure the energy spectra of multi-MeV ion species generated in laser–plasma interactions. The spectrometer features a flexible and reconfigurable design, with modular components tailored for easy adaptation to various experimental setups and rapid deployment. GEANT4-based optical simulations were employed to investigate several active detection schemes using scintillators, allowing us to evaluate their feasibility and to identify limitations, such as with direct scintillation readout or scintillating fiber bundles. These simulations informed the design choices and highlighted the need for continued optimization. Although experimental validation under real conditions remains to be performed, this work lays the foundation for high-repetition-rate, active ion detection compatible with current and upcoming high-intensity laser facilities.
\end{abstract}


\maketitle

\section{\label{Intro}Introduction}

For many years, substantial efforts have been made to study laser–matter interactions and to produce sources of secondary radiations such as energetic ions. The introduction of Chirped Pulse Amplification (CPA) has significantly increased the achievable intensities of the laser systems\cite{STRICKLAND1985219} and contemporary laser facilities are capable of achieving intensities \cite{Bahk:04} at the laser focus of up to \(10^{22} \, \text{W/cm}^2\). During the interaction of such an intense laser pulse with a target, the intensity in its rising edge and/or ASE (Amplified Spontaneous Emission) is powerful enough to convert the material into a plasma. Consequently, the main part of the pulse interacts with a highly ionized and heated plasma which results in particles acceleration of electrons and ions.

However, using solid targets necessitates precise alignment for each shot and results in debris generation and deposition on nearby high-intensity laser optics. This makes it challenging to use in new high-repetition-rate (around 10 Hz and with kHz expected), ultra-high-intensity laser (sources of more intense ElectroMagnetic Pulses) facilities such as Apollon\cite{papadopoulos2016}, ELI\cite{Margarone2018,Radier} or VEGA\cite{Huault2017}. To address these challenges, researchers are exploring self-regenerating targets, including water jets\cite{George_Morrison, Puyuelo-Valdes_2022} or droplets\cite{Karsch, Ter-avetisyan}, cryogenic ribbons\cite{Margarone2016}, tape target\cite{Ehret_2024} and liquid crystal films\cite{Thoss, Poole_2014}. Additionally, near-critical density targets like high-density gas jets are being investigated\cite{Palmer_2011, Haberger_2012, Henares_2019}.

Diagnosing and characterizing the spectra of individual ion species is crucial for understanding the fundamental acceleration mechanisms, that range from Target Normal Sheath Acceleration (TNSA)\cite{TNSA} to Radiation Pressure Acceleration (RPA)\cite{RPA} and Magnetic Vortex Acceleration\cite{Magnetic_Vortex} or Collisionless Shock Acceleration (CSA)\cite{CSA} depending on the target. At high repetition rate, Time-Of-Flight (TOF) method based on a stacked diamond detector structure can be used to measure the velocity of the accelerated particles\cite{TOF}. However, it is not possible with this method to determine the type of particles detected. In addition, this diagnosis is not necessarily ideal given its experimental footprint. Over diagnostic tools frequently employed in single shot - such as Radiochromic film stack detectors\cite{RCF} or nuclear activation methods\cite{Tarisien_2006}  - Thomson Parabola (TP) spectrometers stand out. They provide the ability to separate different ion species by energy and mass-to-charge ratio (A/Z) using static electric and magnetic fields.

\sloppy
A significant drawback of traditional TP spectrometers is their dependence on localization detection system like Imaging Plates (IP) and CR-39\cite{TP1, TP2, TP3}, which require substantial time for removal from the vacuum chamber and subsequent scanning. This limitation hinders their ability to operate at the high repetition rates that modern petawatt (PW) laser systems\cite{Wang:17}, capable of operating at hertz frequencies, can achieve. Micro-channel plates (MCPs) coupled to phosphor screens provide an alternative for rapid data acquisition\cite{MCP1, MCP2, MCP3}; however, they require a high vacuum environment, are costly, and can be easily damaged. All these reasons led us to test the scintillator solution as a particle detection medium, notably because the very fast response time of the latter allows data to be acquired at high repetition rates.

In this study, we propose the design of a Thomson parabola (Sec.~\ref{DevTP}) capable of detecting particles generated by high repetition rate lasers. To achieve this, we will present the results of simulations conducted with various active TP configurations, based on scintillators read by a CMOS camera\cite{activeTP, Exp_CSI} (Sec.~\ref{Active_TP}). An experimental characterization of the different scintillators considered in this work is available in Appendix~\ref{Tests_AIFIRA}.

\section{\label{DevTP} Development of a Thomson Parabola}

\subsection{\label{} Principle and design}
\begin{figure}[h]
\includegraphics[width=0.49\textwidth]{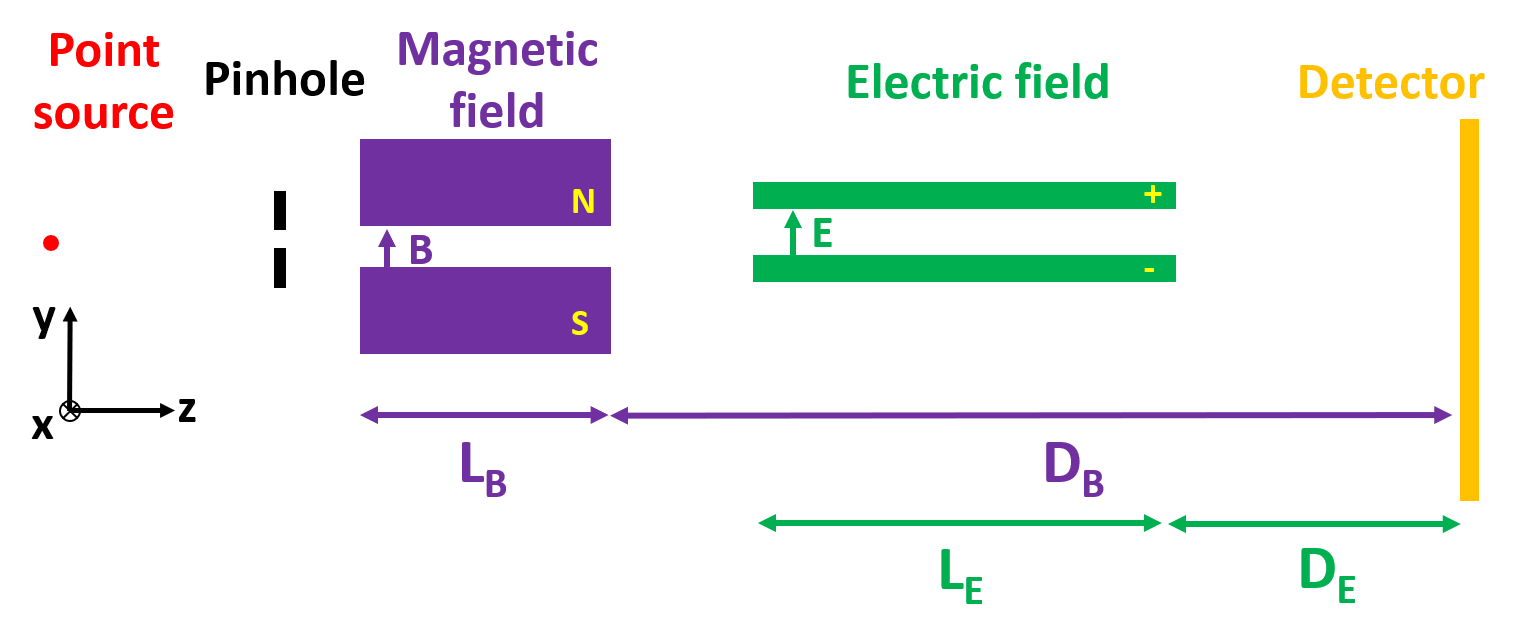}
\caption{\label{fig1} Schematic of a Thomson Parabola spectrometer.}
\end{figure}

One of the most common tools for probing a laser driven ion beam is the Thomson parabola spectrometer (TP). The spectrometer utilizes parallel magnetic and electric fields to sort the different ions species by their velocity and charge-to-mass ratio with electric field deflections in -y direction and magnetic deflections in +x direction according to the Fig.~\ref{fig1}. 

In order to increase the detection solid angle, it is preferable to place the spectrometer as close to the interaction point as possible. However, this creates specific challenges. Firstly, the strong X-ray flux produced by the laser-target interaction can cause significant background on standard detectors, like Imaging Plates (IPs) which decreases the spectrometer's sensitivity unless adequate shielding is used. Secondly, intense Electromagnetic Pulses (EMPs) in the radio-frequency microwave range generated during the interaction can distort the deflecting fields, potentially causing modulation and overlap of the expected traces, affecting the accuracy of particle discrimination\cite{Distorsion_EMP}. These challenges become more significant as the spectrometer is positioned closer to the target. Meanwhile, for most spectrometers, electrostatic and magnetostatic deflections are performed in separate stages leading to relatively long devices, placing a large structure inside the vacuum chamber near the target can be difficult due to the presence of various other components in most experimental setups.

Our TP design presents three distinct parts:
\begin{itemize}
    \item The first part includes the magnetic and electric components, as well as the entrance pinhole.  
    \item The second part, called the "free-flight" section, does not contain any elements but allows particles to continue propagating in straight lines.  
    \item The third part includes the localization detector, which can be adapted to various detection configurations (IPs, MCP, scintillator, etc.).
\end{itemize}

\begin{figure}[h]
\includegraphics[width=0.5\textwidth]{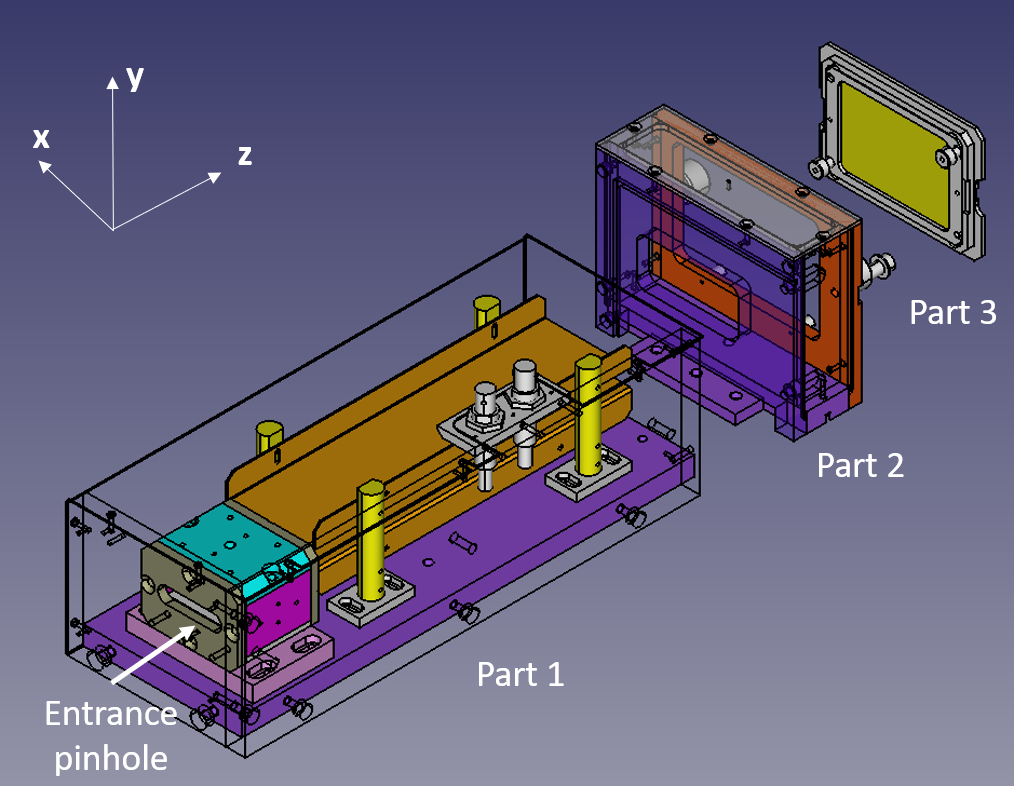}
\caption{\label{fig2} Exploded view of the different parts of the TP.}
\end{figure}

This mechanical configuration (visible on Fig.~\ref{fig2}) allows for easy switching between different setups by removing parts 2 and/or 3. 

In the following sections, we will present in details the elements of the TP.

\subsubsection{\label{} Magnetic field part}\mbox{}\\

A design criterion for the layout of the magnet was the use of readily available, relatively inexpensive cuboid-shaped Nd magnets. Such magnets can be found “off-the-shelf” at many suppliers. Two such magnets facing each other produce a dipole magnetic field in the gap between. Provided the gap width d is small compared to the magnet length L$_B$ and width W, the magnetic field is homogenous over most of the field volume. In order to achieve sufficient deflection up to the highest energies, both the magnetic field strength and its length L$_B$ have to be as large as reasonably achievable. Furthermore, the width has to be large enough to still cover the trajectories of particles with the lowest energy to be measured (strongest deflection in magnetic field). Given the thickness T and the remanence field Br, the field in the center of the gap can be calculated using\cite{zielbauer_2022_7305167}:
\begin{equation}
\begin{split}
    B_{center} = \frac{2}{\pi} B_r \left[ \tan^{-1} \left(\frac{L_B \times W}{d \sqrt{d^2 + L_B^2 + W^2}}\right) \right. \\
    \left. - \tan^{-1} \left(\frac{L_B \times W}{(d + 4T) \sqrt{(d+4T)^2 + L_B^2 + W^2}}\right) \right]   
\end{split}
\end{equation}

\begin{figure}[h]
\centering
\includegraphics[width=0.35\textwidth]{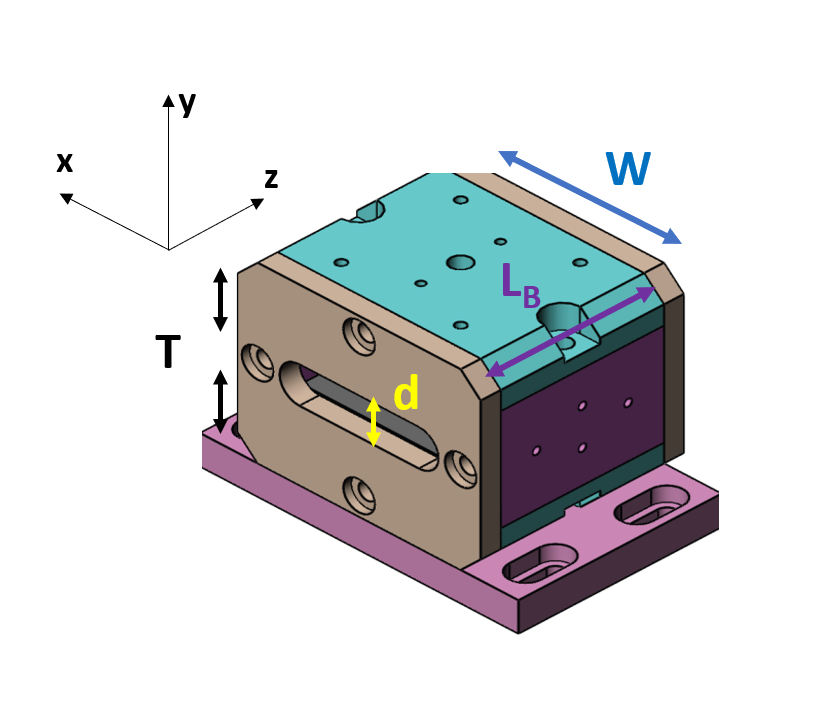}
\caption{\label{figyoke} Yoke plan used to limit magnetic field flux leaks.}
\end{figure}

We propose using two magnets with dimensions of 40 mm in length (L$_B$), 40 mm in width (W), and 10 mm in thickness (T), and a remanent magnetization (Br) of 1.27 T (magnetization grade N40) as it can be seen on Fig.~\ref{figyoke}. With a gap width (d) of 5 mm, the above equation gives a magnetic field strength B$_{center}$ of 0.76 T. This gap width offers a good compromise between alignment simplicity and minimizing field inhomogeneities near the edges of the magnets.

The magnets are incorporated within an iron yoke to create a closed magnetic circuit and to direct the magnetic field lines outside of the gap. The yoke's cross-sectional area (Fig.~\ref{figyoke}) is designed to be sufficiently large to remain below the saturation magnetization of typical high-permeability ferromagnetic materials (such as soft iron). This helps to minimize magnetic leakage flux and stray fields outside of the yoke.

\begin{figure}[h]
\includegraphics[width=0.5\textwidth]{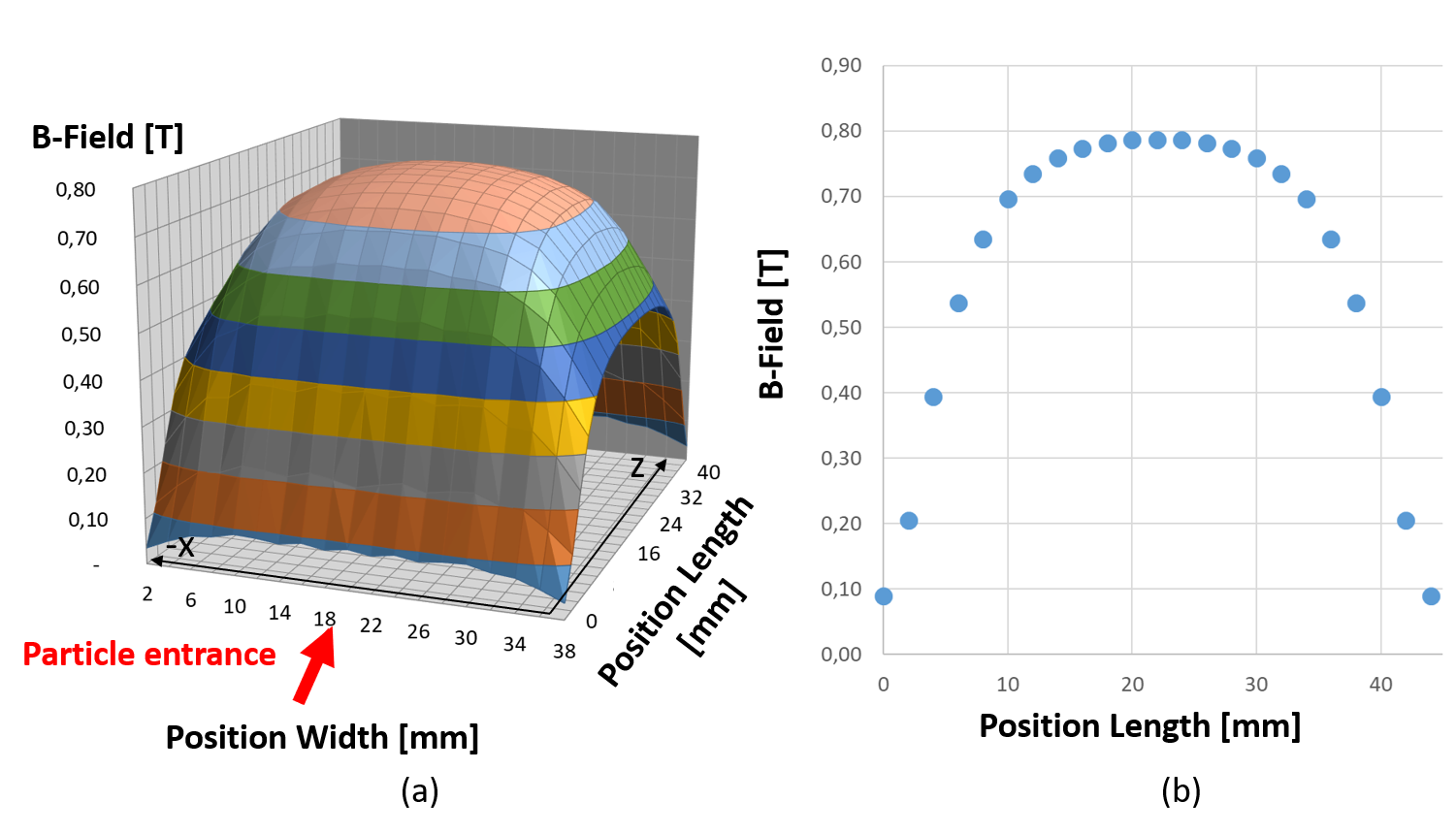}
\caption{\label{fig4} Magnetic field measurements data. (a) corresponds to 3D surface plot of the vertical component of the magnetic field and for (b) line out of this component along the particle entrance axis.}
\end{figure}

A measurement of the magnetic field was performed and recorded by inserting a Hall probe along the ion's path in the magnetic part (Fig.~\ref{fig4}). The maximum value of magnetic field is 0.78 T which is in agreement with our estimation. At the exit of the magnetic field, the maximum x deflection for protons reaching the detector is calculated to be 4 mm which corresponds to the lowest proton energy measurable of 1.2 MeV. This is still well within the width of the homogeneous magnetic field (W = 40 mm).

\subsubsection{\label{} Electric field part}\mbox{}\\

The electric field introduces separation in the direction perpendicular to the dispersion caused by the magnetic field. Since the electrostatic force is proportional to \(q/m\), this leads to a splitting of traces on the detector plane for different particle species. To simplify manufacturing, we have chosen two parallel capacitor plates. At a distance of \(d_E\), the electric field between the plates with an applied voltage \(U\) is given by \(E = U/d_E\). 

\begin{figure}[h]
\centering
\includegraphics[width=0.48\textwidth]{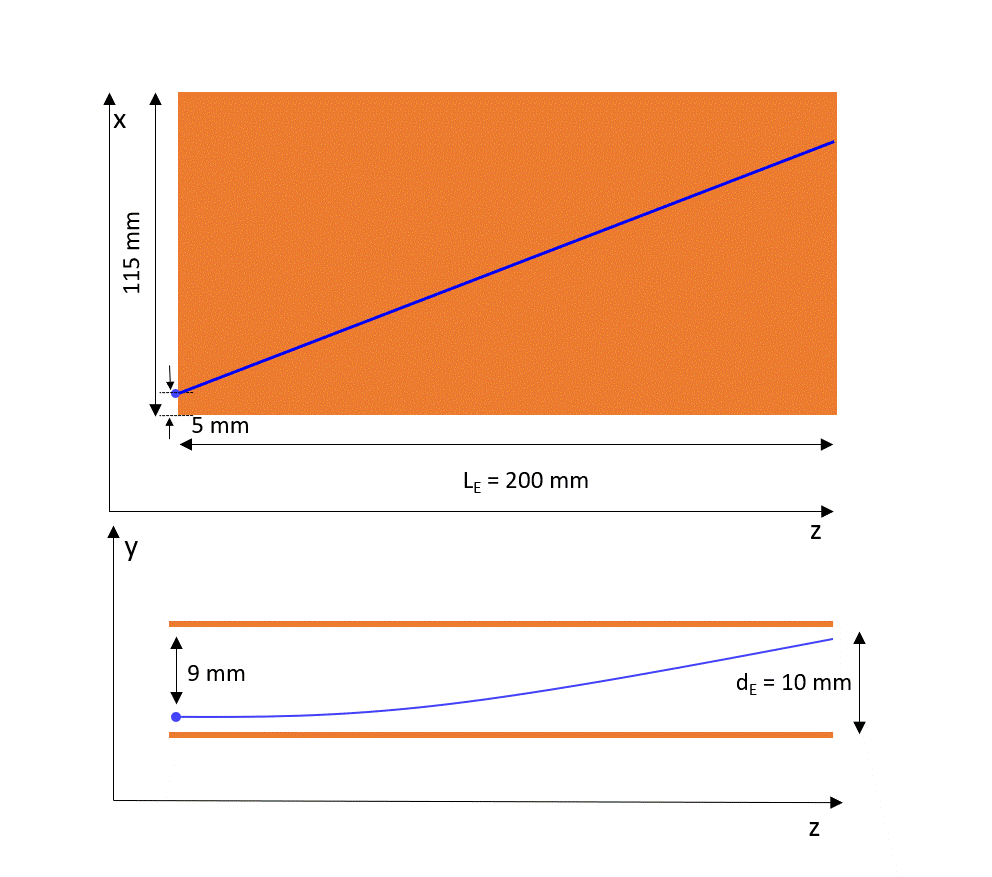}
\caption{\label{fig5} Dimensions of the electric field plates where the blue lines indicate the beam trajectory of 1.2 MeV protons in the capacitor.}
\end{figure}

Our baseline design (see Fig.~\ref{fig5}) features plates with a length of \(L_E = 200\) mm (to provide adequate separation between traces of different species), a width of 115 mm (to cover particles dispersed by the magnetic field down to 1.2 MeV), and a separation of \(d_E = 10\) mm. The system is expected to operate with a voltage of \(\pm 5\) kV applied to the plates, a configuration found to be reliable with standard coaxial connectors (SHV, BNC-HV). The beam enters the capacitor at a distance y=1 mm with respect to the bottom plate in order to maximise the y deviation due to the electric field and of x=5 mm from the edge of plates keeping the straight line trajectory in the homogenous electric field area. For manufacturing reasons, field free regions of 10 mm are foreseen between the rear end of the magnet and front of the capacitor, and 6 mm between the rear end of the capacitor and the structure of the second part of the TP.

\subsubsection{\label{} Detectors and alignment laser support}\mbox{}\\

The charged particles entering the spectrometer from the pinhole, are deflected by the magnectic and electric fields, cross the free flight zone and reach the localization detector. It is expected to have different mounts according to the type of detectors used (IPs, MCP, Scintillator, CMOS, ...). A laser diode can be temporarily fixed to the rear of the spectrometer housing, with its beam entering through a small aperture. When the laser beam is precisely aligned to traverse both the entrance and exit holes, it defines the trajectory of photons entering the spectrometer. This alignment allows the spectrometer to be accurately pointed towards the source, such as the target under investigation.

To ensure more flexibility concerning the detector part, the system has been designed like cartridges that can be inserted into the third structure. This allows for quick switching from one detector to another without having to disassemble the corresponding structure.

\subsection{\label{} Simulation and analysis of the spectrometer performance}

Several factors determine the TP's energy resolution ($\frac{\Delta E}{E}$), including the detector's spatial resolution, the maximum dispersion, the aperture diameter at the entrance of the TP, and the divergence of the beam. The TP's dispersion depends on the strength of the electric and magnetic fields, the charge-to-mass ratio as well as the kinetic energy of the deflected ions. Simulations of the entire spectrometer conducted with GEANT4 \cite{GEANT4} allowed us to establish a dispersion relation for each charge-to-mass ratio within a specific energy range. The primary factor limiting the TP's energy resolution is the entrance pinhole solid angle ($<$ µsr). Thus, in all our studies, the pinhole diameter at the entrance of the TP (30 cm from the target) was set to 100 $\mu$m in the simulations. We considered in the following IP detectors with a spatial resolution of 50 $\mu$m (in the most optimistic scenario) to isolate the effects of other parameters.

In our case, the species separation can be increased by three possible methods: 

\begin{itemize}
    \item Increasing D$_E$,
    \item increasing the electric field strength,
    \item increasing L$_E$.
\end{itemize}

Increasing \( D_E \) significantly in order to improve the resolution is generally not a feasible option for several reasons. For instance, placing a spectrometer with a large \( D_E \) inside a compact interaction chamber while maintaining a reasonable distance from the target can be challenging. Moreover, moving the detector plane much further away from the end of the electric field plates will proportionally increase the magnetic field dispersion (i.e., \( D_B \)). This may be undesirable given the limited size of the detector. Additionally, any modification that increases the dispersion of ion traces will reduce the surface density of the particle beam reaching the detector plane, resulting in a lower signal-to-noise ratio.

The strength of the electric field can be enhanced either by reducing the distance between the electrodes or by increasing the applied voltage. However, as seen before, the main limitation is the risk of breakdown. Based on experimental expertise, electric fields up to approximately \(2 \times 10^6\) V/m can be applied in a typical experimental chamber at vacuum pressures ranging from \(10^{-4}\) to \(10^{-5}\) mbar. Conversely, there are no fundamental limitations to increasing the length of the electric plates except for losing low energy protons that can interact with too long plates.

\begin{figure}[h]
\includegraphics[width=0.5\textwidth]{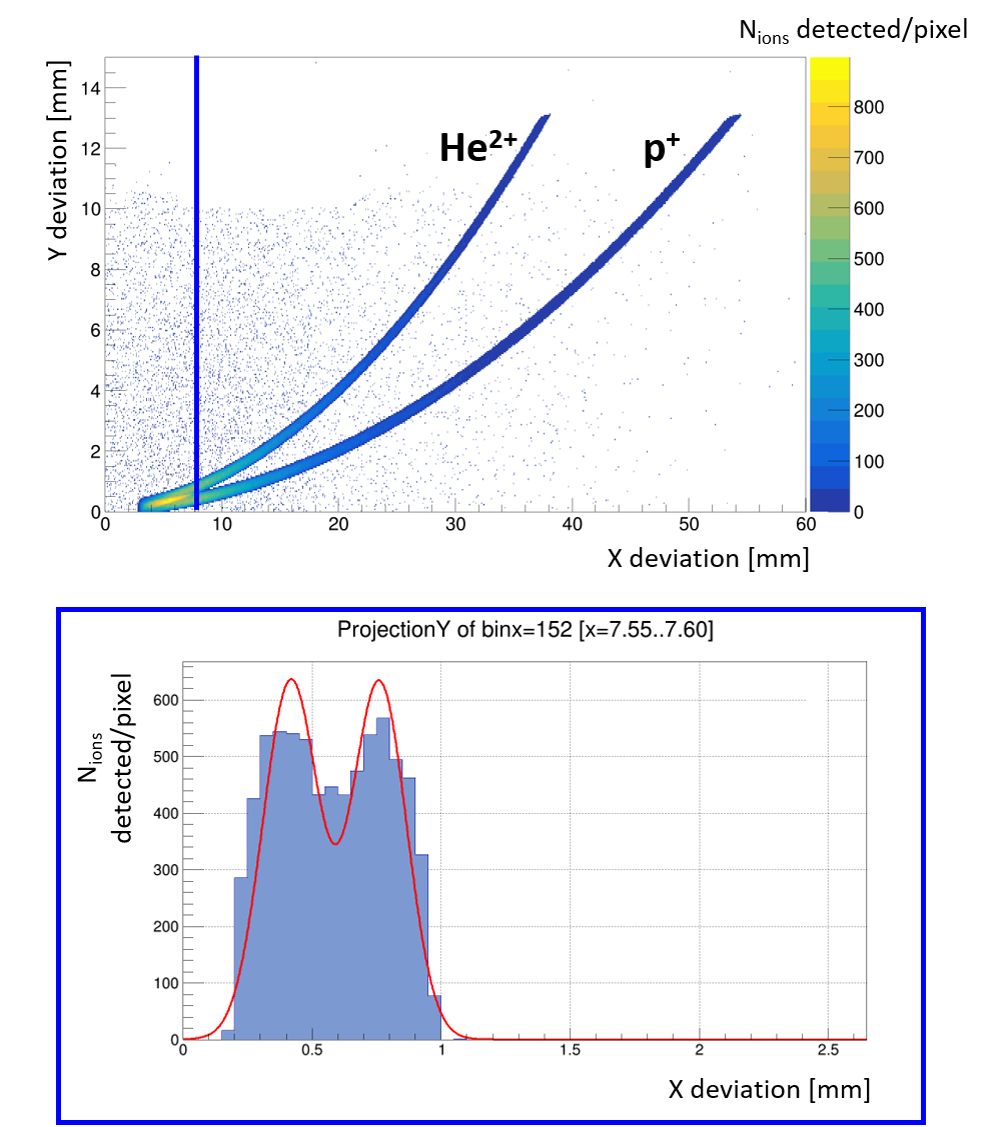}
\caption{\label{fig6}Top: 2D distribution of particle trajectories (X vs. Y deviation in mm), showing clear separation between He$^{2+}$ and p$^+$ ions. The color scale indicates event density. The vertical blue line marks the last X-bin for which two distinct peaks are resolved in the Y-projection, as determined by a ROOT peak-finding algorithm.\\
Bottom: Y-projection histogram at the selected X-bin. Two peaks correspond to He$^{2+}$ and p$^+$. The red line shows a double-Gaussian fit used for species discrimination.}
\end{figure}

The resolution power, which enables the discrimination of a p$^+$ trace from a He$^{2+}$ trace, was determined using data from GEANT4 simulations and a post-analysis performed with the ROOT software\cite{Root}. For this purpose, at each position on the magnetic deflection axis (i.e., particle energy), we checked whether an algorithm for peak detection could still distinguish between the two contributions on the electric deflection axis (Fig.~\ref{fig6}). If this condition was not met, we referred to the position on the magnetic axis to determine the maximum proton energy that could be discriminated from He$^{2+}$ ions.

A large number of simulations were carried out to test various configurations. These simulations led us to the following optimal configuration:
\begin{itemize}
    \item Pinhole radius = 100 µm  
    \item \(L_B = 40\) mm with \(B = 0.76\) T  
    \item \(L_E = 200\) mm with \(E = 1000\) kV/m  
    \item \(D_B = 265\) mm  
    \item \(D_E = 50\) mm  
    \item Total length = 300 mm  
\end{itemize}

This configuration allows us to achieve the following performances :  
\begin{itemize}
    \item Energy resolution for p$^+$ @ 10 (100) MeV = 0.5 (1.7)$\%$
    \item p$^+$/He$^{2+}$ discrimination possible up to 26 MeV  
    \item Minimum observable p$^+$ energy around 1 MeV  
    \item Maximum dispersion at the detector plane of 56.2 mm  
\end{itemize}

\subsection{\label{} Experimental tests with IPs at CLPU}

Our Thomson Parabola was tested for TNSA shots equiped with IP as localisation detector during a campaign on Vega~3 at CLPU\cite{Marine_VEGA3}. The Ti:Sa laser VEGA-3 can deliver 30 J in a pulse duration of 30 fs at a wavelength of 800 nm. The laser spot diameter is about 11 $\mu$m FWHM with 25$\%$ of the energy inside the central peak. During the experiment, the pulse duration was adjusted to 200 fs, and intensity on target was $\approx$ 3.5 × 10$^{19}$ W/cm$^2$. A 200 $\mu$m diameter pinhole was used at the entrance of the TP, and a TR type Imaging Plate was used for the detection.

\begin{figure}[h]
\includegraphics[width=0.53\textwidth]{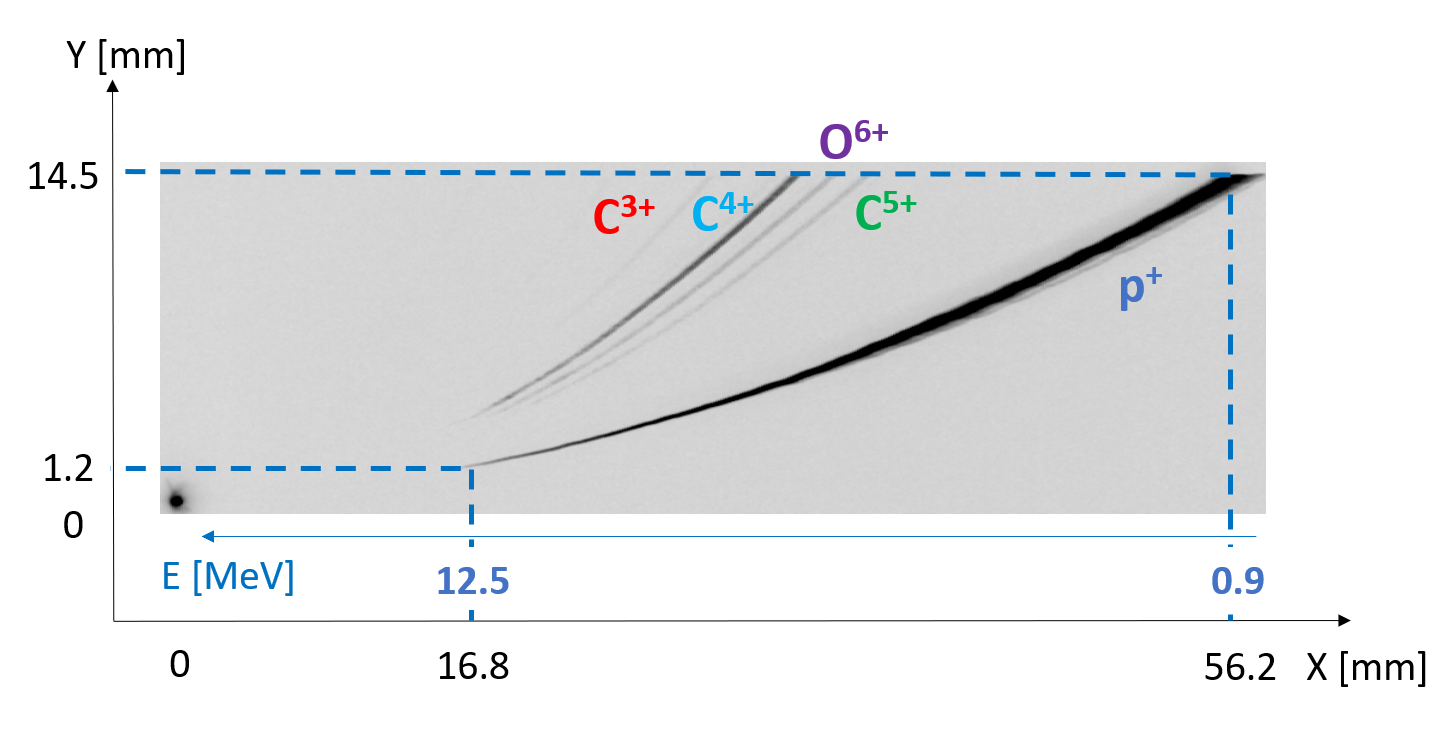}
\caption{\label{fig7}Averaged on 50 shots (aluminium 8 $\mu$m targets) Imaging plate scan on Vega 3.}
\end{figure}

The protons and ions (C$^{3+}$, C$^{4+}$,O$^{6+}$ and C$^{5+}$) were clearly detected with a minimum proton threshold energy of 0.9 MeV (see Fig.~\ref{fig7}). The ions parabola were in agreement with predicted simulations using a map of B field and no significant EMP perturbations were observed. The thickness of the trace is linked to the size and conical shape of the current pinhole.

\section{\label{Active_TP} Development of an Active Thomson Parabola}

We have explored various configurations of detection systems (part 3) based on scintillators. 

The following sections present a detailed analysis of these configurations, highlighting their performance in terms of photon detection, energy resolution, and practical implementation challenges. By comparing the number of photons collected and the energy resolution achieved, we aim to identify the most feasible and efficient setup for our experimental needs. Additionally, we will examine in Section \ref{Sc-comparison} the impact of different scintillator materials on detection efficiency and consider their responses to X/$\gamma$ radiation, which is crucial for maximizing signal to noise ratio and ensuring accurate ion detection.

This comprehensive evaluation will guide our selection of the optimal configuration for future experiments, ensuring robust and reliable performance under varying conditions.

Based on this framework, several ideas emerged to adapt our TP:
\begin{itemize}
    \item Use a scintillator in the detection plane with the camera in direct contact (Section \ref{Sc-camera-contact}).
    \item Use a network of scintillating optical fibers in the detection plane to enable remote observation by the camera (Section \ref{Fibers-camera}).
    \item Use a scintillator in the detection plane with remote observation through an optical fiber array (Section \ref{Sc-fibers-camera}).
    \item Use a scintillator in the detection plane with remote observation via an optical imaging system (Section \ref{Sc-lens-camera}).

\end{itemize}

To test and compare the performance of these different solutions, optical simulations were conducted using GEANT4, which will be briefly presented in the next section. 

\subsection{\label{} Optical Simulation with GEANT4}
\sloppy
When the wavelength of a photon is much greater than the atomic spacing (typically $\lambda$ = 10 nm), it can be considered as an optical photon. With GEANT4, light can be treated according to this definition (G4OpticalPhoton) independently from higher energy gamma radiation (G4Gamma). The two classes are distinct, and there are no possible transitions based on the energy of the photon. It is therefore essential to clearly define the type of photon used according to the considered energy range. Furthermore, although described in its wave aspect, the information on the phase of the photon is not managed by the simulation, making it impossible to generate potential interferences. However, such phenomena are very marginal in the case of scintillators with emission spectra spanning over several hundred nanometers.

The different processes that can apply to optical photons are as follows:

\begin{itemize}
    \item light production by scintillation (G4Scintillation), by Cherenkov emission (G4Cherenkov), or by transition radiation (G4TransitionRadiation),
    \item light attenuation during its propagation (G4OpAbsorption),
absorption/re-emission phenomena with wavelength shifting of photons by secondary agents (G4OpWLS),
    \item Rayleigh scattering (G4OpRayleigh),
    \item various possible interactions such as reflection, absorption, and transmission at the interfaces between two media (G4OpBoundary).
\end{itemize}

All these processes are known in their mathematical form. The optical path of light is defined by Fermat's principle, the behavior at optical interfaces is described by Snell's law, light attenuation in a medium is described by the Beer-Lambert law, and Cherenkov radiation is also generated. These processes depend on various parameters (optical index, absorption length, reflectivity index) that each have a dependency on the wavelength of the optical photons. All these parameters must be provided as input to the simulation. For this purpose, GEANT4 allows the creation of property tables (G4MaterialPropertiesTable) in the form of arrays of values depending on the wavelength. More details on the different properties used by this kind of simulations are given in this publication\cite{huber}.

\begin{figure}[h]
\includegraphics[width=0.51\textwidth]{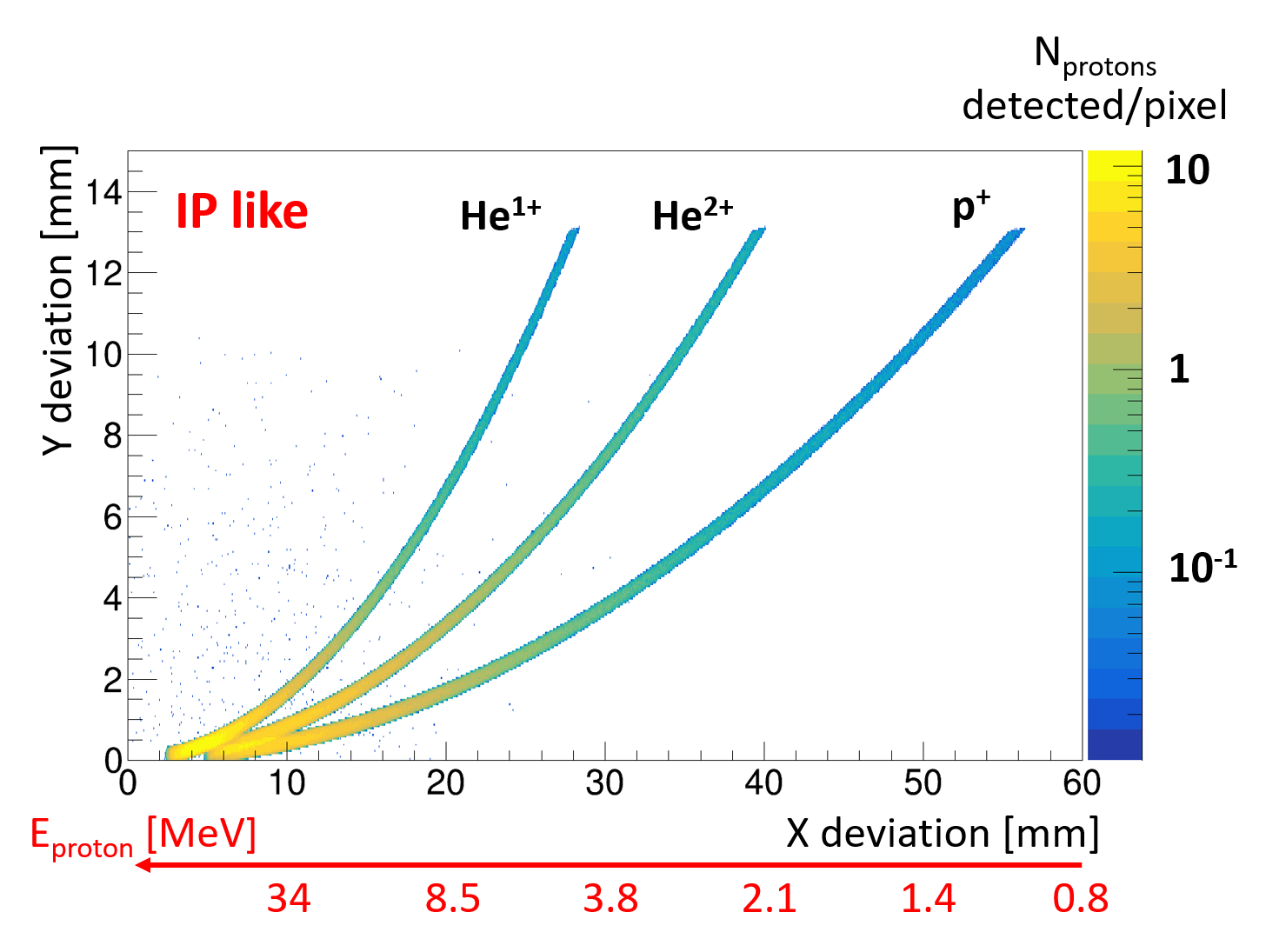}
\caption{\label{figIP}Simulations of p$^{+}$, He$^{1+}$ and He$^{2+}$ in the IP like configuration with a pixel size of 50 µm.}
\end{figure}

In order to compare the performances of our 4 configurations, we simulated 3 different ions (p$^+$, He$^+$ and He$^{2+}$) according to a TNSA like profile with 3 parameters : cutoff energy (110 MeV), temperature (E$_0$ = 30 MeV) and different number of ions generated at a distance of 30 cm from the TP's pinhole in a solid angle covering this latter. Fig.~\ref{figIP} illustrates a typical example of the parabolas corresponding to these three ions when using an IP. In this configuration, we are operating under the most optimistic conditions achievable with our setup, which will serve as a reference point to assess the impact of configuration changes on the detection of these ions. 

Concerning the detection, we chose to work with a CMOS Camera from HAMAMATSU (Orca-Flash 4.0 LT Plus) with a pitch of 6.5 $\mu$m and a maximum Quantum Efficiency of 82 $\%$ at 560 nm. All of that will enable us to distinguish the specific effects of each parameter (configuration type, scintillator type, scintillator thickness) on the performance of the TP, such as the signal to noise ratio on localisation detector, the width of the traces, their possible discrimination, and thus the associated energy resolution.

For each configuration involving a scintillator, two different thicknesses (0.1 and 1 mm) of plastic scintillator (EJ-262) will be tested to assess the impact of this parameter on the results. 

\begin{figure}[h]
\includegraphics[width=0.51\textwidth]{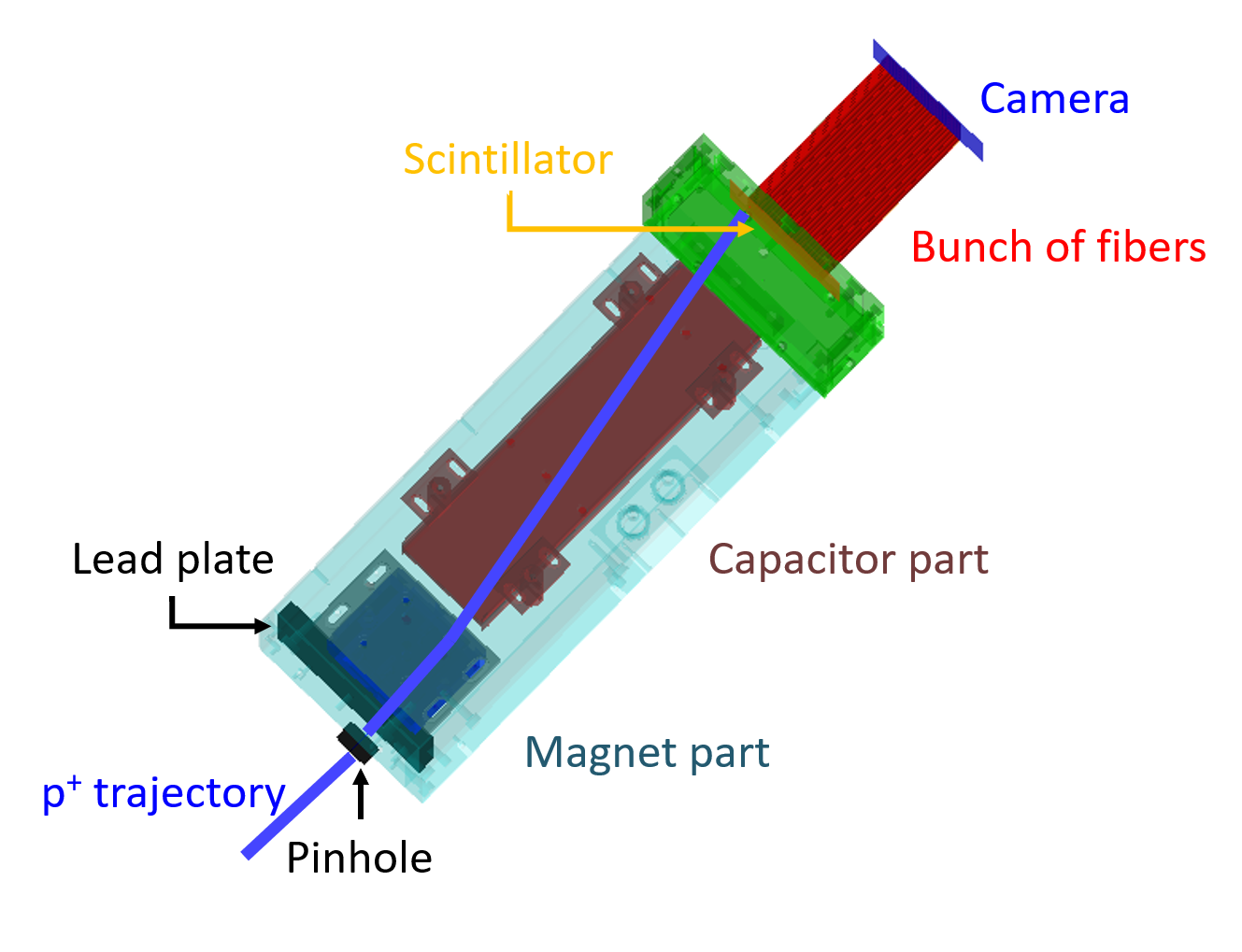}
\caption{\label{figfibersGEANT4}Trajectory of a 2 MeV proton through the TP in GEANT4 in the case where a scintillator is in the detection with a remote observation via an optical fiber array.}
\end{figure}

For configurations involving the fiber array, the fiber size has been set to a diameter of 200 µm, which corresponds to the minimum size of scintillating fibers sold by our supplier, Kuraray. The properties of these fibers can be found on the Kuraray's website\cite{Kuraray} where the scintillator material corresponds to an EJ-212 scintillator (see Table~\ref{tab:1} in Section~\ref{Sc-comparison}). The different results presented will correspond to the Round Fiber Multi-Cladding solution. As an illustration, Fig.~\ref{figfibersGEANT4} shows the simulated geometry in GEANT4 when a fiber network is used to guide the scintillator's light to the camera.

For clarity purposes, we will first present the results specific to each configuration for a given scintillator (EJ-262) and the two tested thicknesses in order to better identify the sensitivity of the TP with these parameters (Sections \ref{Sc-camera-contact}, \ref{Fibers-camera}, \ref{Sc-fibers-camera} and \ref{Sc-lens-camera}). Following these sections, we will attempt to compare and summarize what these simulations can teach us (Section \ref{Configuration-resume}), before concluding with the impact on the results when changing the type of scintillator (Section\ref{Sc-comparison}).

\subsection{\label{Sc-camera-contact} Scintillator with camera in contact}

\begin{figure}[h]
\includegraphics[width=0.51\textwidth]{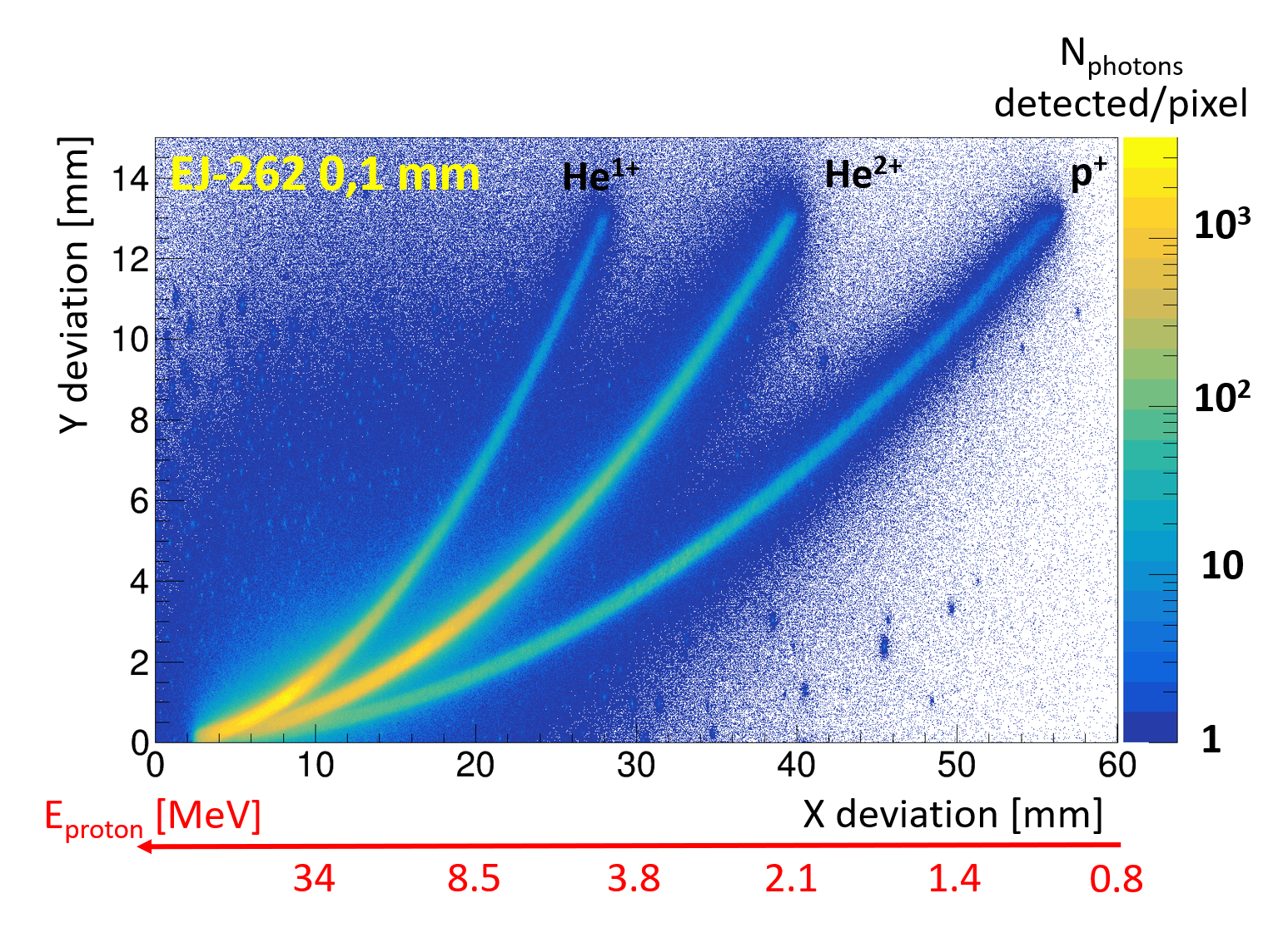}
\caption{\label{figSccontact0.1}Optical simulations of p$^{+}$, He$^{1+}$ and He$^{2+}$ in the configuration with the camera in direct contact with a 0.1 mm EJ-262 scintillator. The binning corresponds to the pitch of our camera (6.5 µm).}
\end{figure}

\begin{figure}[h]
\includegraphics[width=0.51\textwidth]{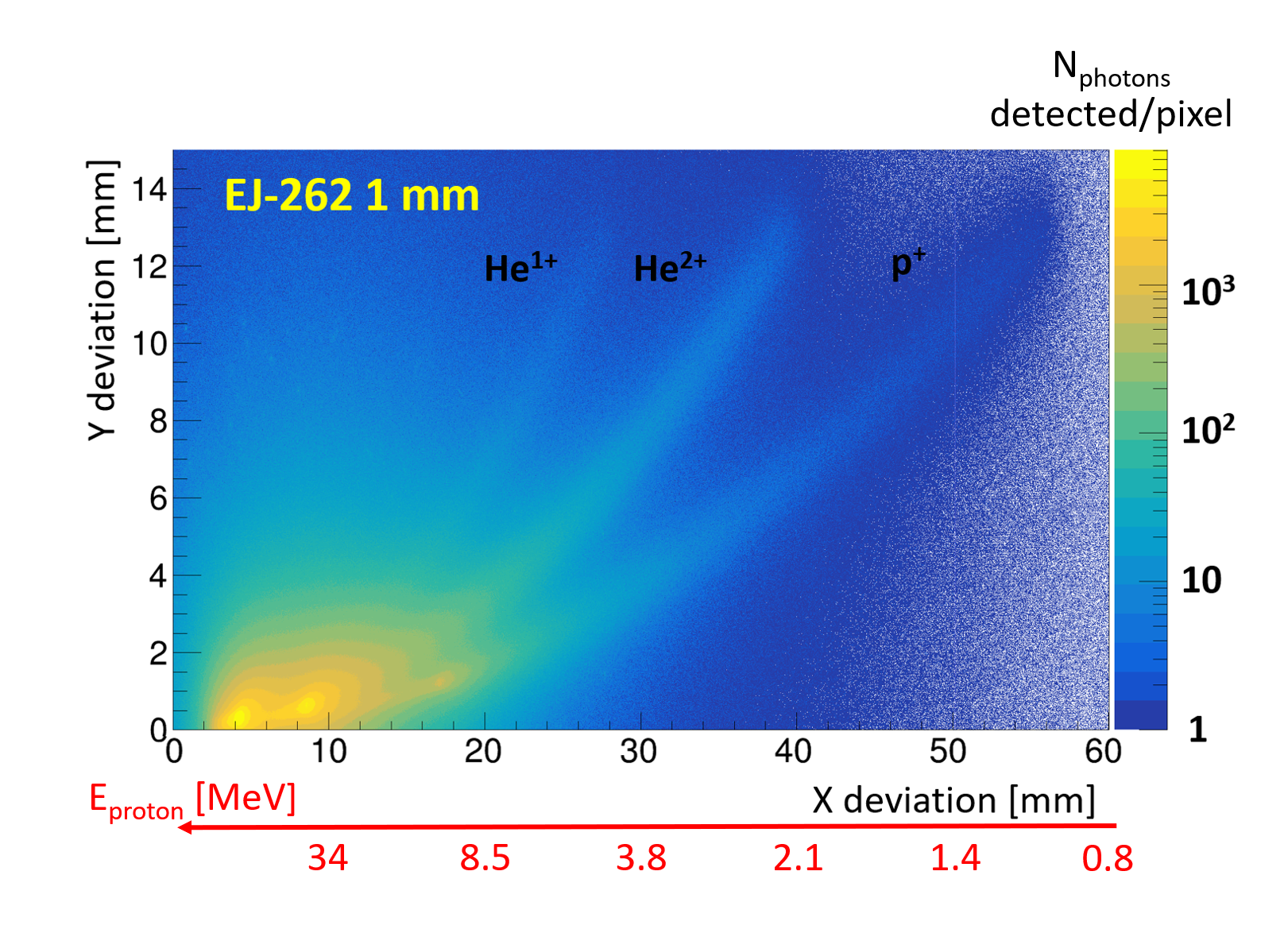}
\caption{\label{figSccontact1}Optical simulation of p$^{+}$, He$^{1+}$ and He$^{2+}$ in the configuration with the camera in direct contact with a 1 mm EJ-262 scintillator. The binning corresponds to the pitch of our camera (6.5 µm).}
\end{figure}

This configuration, the simplest to set up, allows for optimal optical performance while minimizing space requirements. The simulation results are shown in Fig.~\ref{figSccontact0.1} and Fig.~\ref{figSccontact1} which correspond to scintillators of 0.1 mm and 1 mm respectively.

The most striking observation is the significant increase in background noise across the entire camera compared to detection via IP. This is not surprising, however, given that scintillation is emitted within a 4$\pi$ solid angle, and the proximity of the camera to the scintillator results in the near-total detection of all produced photons from the scintillator making the trace broader. This effect becomes even more pronounced when the scintillator is thicker, as the photons have more space to propagate, thereby increasing the dispersion of their detection points relative to their emission points. Indeed, without optical system, the detection plane on the camera will correspond to the exit plane of the scintillator. As a result of this effect, we can see a loss of information about the high energy ions when the scintillator thickness is 1 mm. In addition, the traces tend to thicken for the lowest energies, as the ions are stopped in the first layers of the scintillator, which subsequently increases the dispersion of photons reaching the camera. When this phenomenon is minimized by reducing the thickness of the scintillator, we observe in Fig.~\ref{figSccontact0.1} that access to different information remains possible, despite a slight increase in trace width.

Firstly, there is the need to use thin scintillators to avoid degrading energy resolution (discussion in Section.~\ref{Configuration-resume}). Secondly, if the scintillator is thin, energetic ions could pass through the scintillator and interact with the camera. Beyond the difficulty of evaluating the impact of ions on the camera compared to the detected photon signal, this could primarily result in potential damage to the detector according to the laser power configuration. In a publication, J.Fuchs et al.\cite{Fuchs} report the absolute calibration of CMOS detector with particles created by a 8.1 J per pulse laser. The use of CMOS detectors at higher laser energies such as the ones expected on APOLLON is still to be demonstrated. Thirdly, a compact solution like this can also raise issues of EMP management. 

\subsection{\label{Fibers-camera} Scintillating fibers imaged by a camera}

At first glance, this configuration seemed to be the most promising of all. Indeed, this configuration allows for an optimal transfer of light while enabling the measurement to be moved outside the experimental chamber to avoid issues related to Residual Gas Analysis (RGA) and EMP.

\begin{figure}[h]
\includegraphics[width=0.51\textwidth]{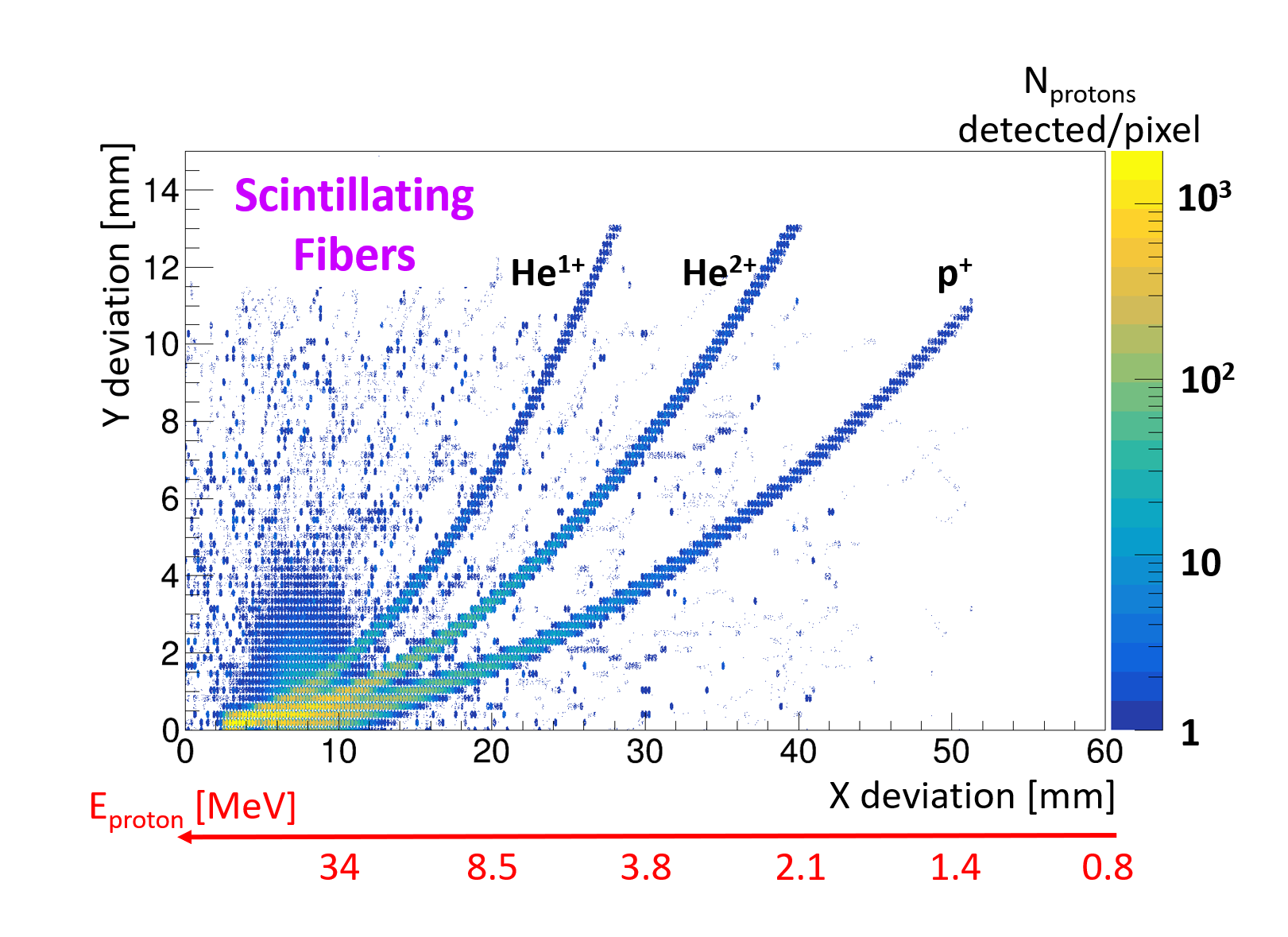}
\caption{\label{fig13}Optical simulations of p$^{+}$, He$^{1+}$ and He$^{2+}$ in the configuration where scintillating fibers (pitch of 200 $\mu$m) are imaged by a camera. The binning corresponds to the pitch of our camera (6.5~µm).}
\end{figure}

However, the simulations of the TP with all the fibers considered (with a 1 m length) quickly revealed a major problem, visible in Fig.~\ref{fig13}. Beyond a certain energy (30 MeV in our example which corresponds to a deviation of $\sim$ 10 mm along x axis), protons are energetic enough to deposit energy in one fiber, pass through it, and then deposit the rest of their energy in one or more other fibers. This results in significant smearing in a region where energy resolution is crucial. Therefore, it could be difficult to use this type of configuration for experiments aiming to measure protons with energies above 30 MeV. However, it is worth noting that this phenomenon occurs within an energy range where discrimination is no longer achievable with IPs, which could ultimately limit the impact of this phenomenon on the obtained results.

Anyway, for experiments not targeting these energy ranges, the energy resolution will not only depend on the cameras's pitch but will also depend on the pitch of the fibers, which currently cannot go below 200 µm due to its design with the scintillating material as the core within the fiber.

\subsection{\label{Sc-fibers-camera} Light from scintillator collected by a fibers network and imaged by a camera}

\begin{figure}[h]
\includegraphics[width=0.51\textwidth]{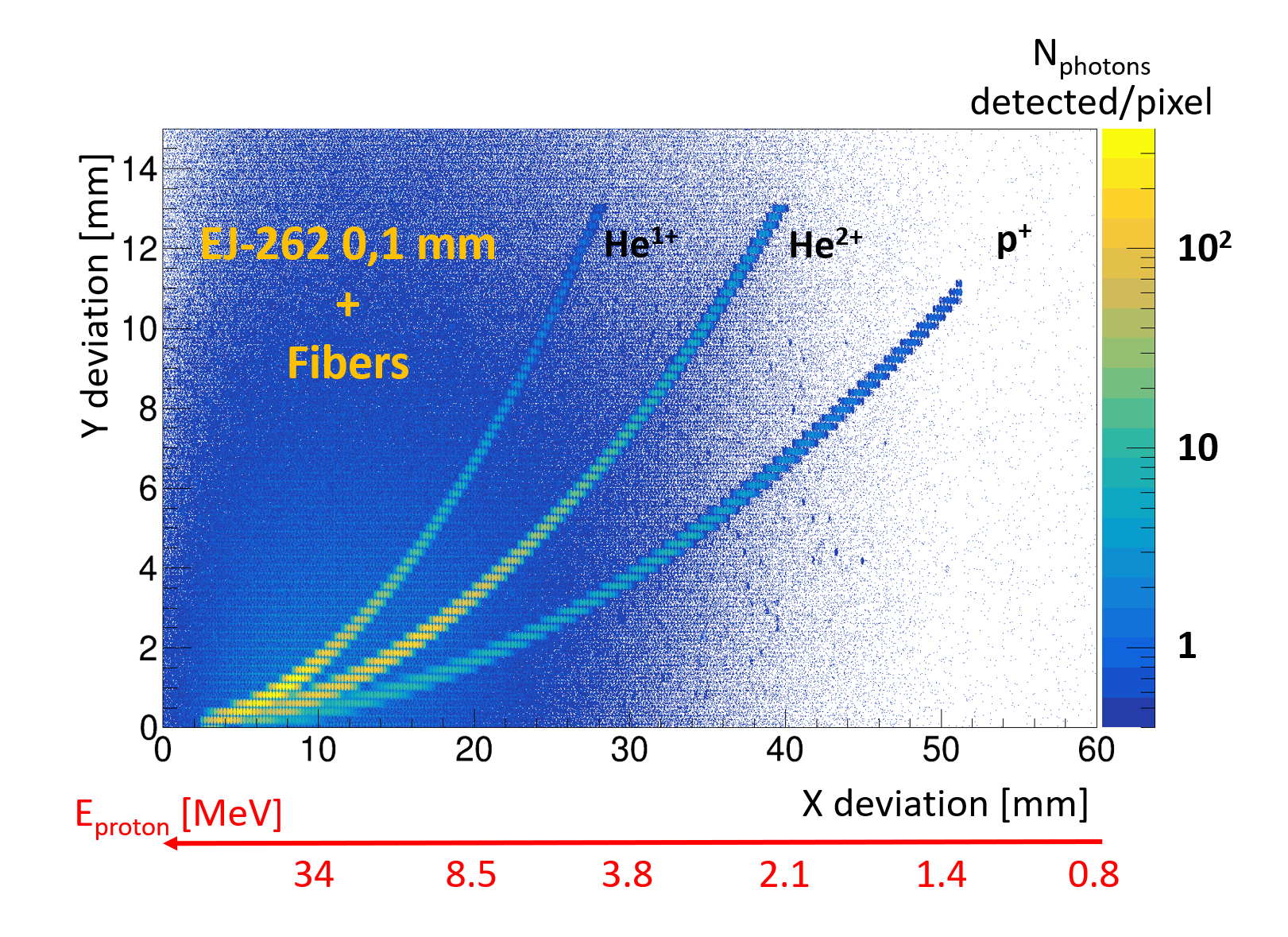}
\caption{\label{figfibers0.1}Optical simulations of p$^{+}$, He$^{1+}$ and He$^{2+}$ in the configuration where the photons emitted by scintillator of 0.1 mm are collected with a bunch of fibers with a pitch of 200 $\mu$m. The binning corresponds to the pitch of our camera (6.5 µm).}
\end{figure}

\begin{figure}[h]
\includegraphics[width=0.52\textwidth]{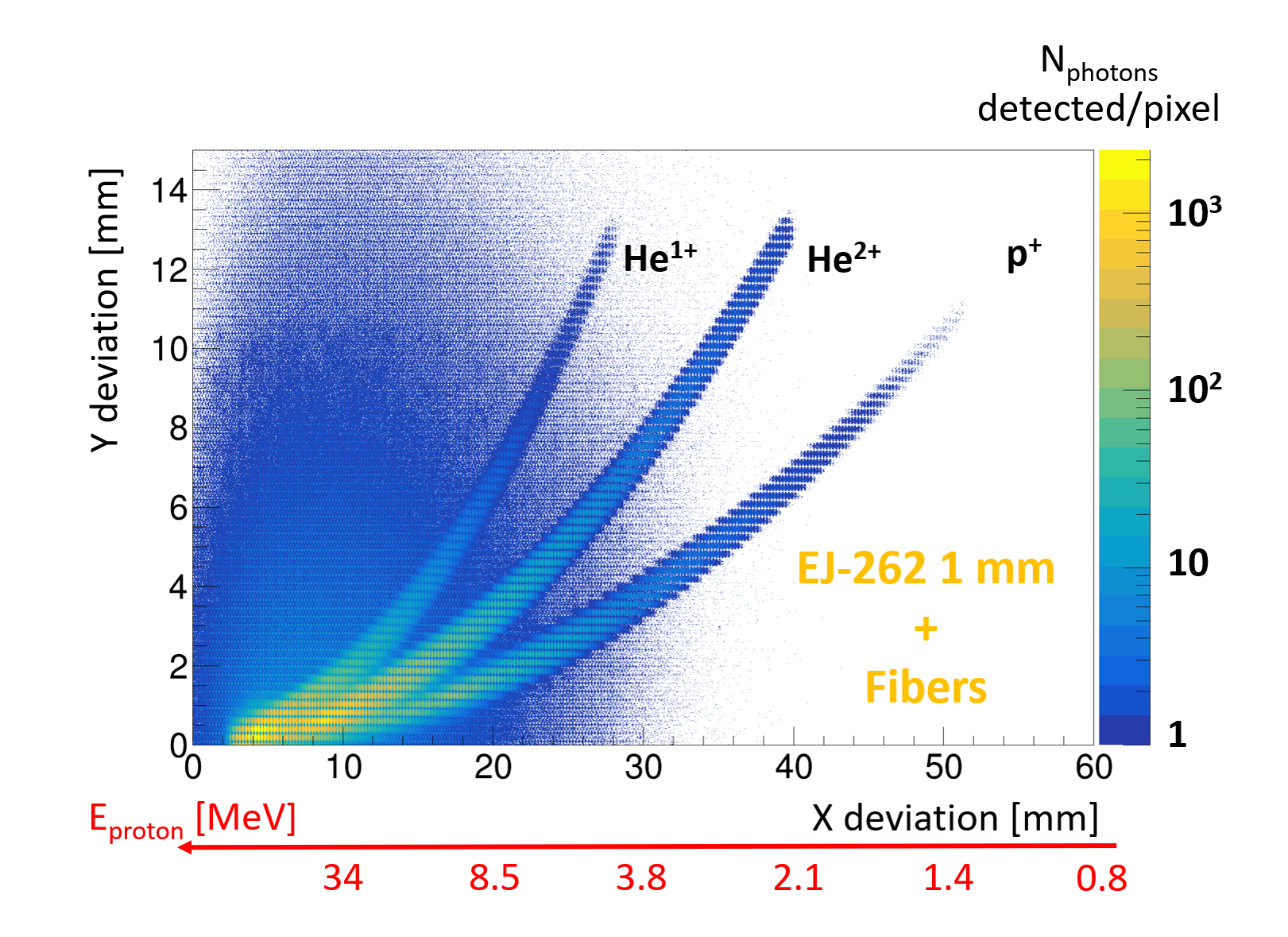}
\caption{\label{figfibers1}Optical simulations of p$^{+}$, He$^{1+}$ and He$^{2+}$ in the configuration where the photons emitted by scintillator of 1 mm are collected with a bunch of fibers with a pitch of 200 $\mu$m. The binning corresponds to the pitch of our camera (6.5 µm).}
\end{figure}

This configuration is similar to the previous one, except that an array of fibers with a 200 µm pitch is installed to create the connection between the scintillator and the camera (see Fig.~\ref{figfibersGEANT4}). The simulation results are shown in Fig.~\ref{figfibers0.1} and Fig.~\ref{figfibers1} which correspond to scintillators of 0.1 mm and 1 mm respectively.

Overall, the same effects as in the first configuration (scintillator and camera in contact) can be observed, which is not surprising given the geometric similarity between the two configurations. However, using the fiber array will necessarily result in fewer photons being detected by the camera compared to the direct contact configuration (factor 10-100) due to the influence of the numerical aperture of the fibers and the loss of some photons during their propagation within the fibers (Fig.~\ref{fignphglobal}). Nevertheless, the numerical aperture of the fibers also helps filter out certain unwanted photons, thereby slightly reducing the background noise on our camera. This makes it possible to increase the thickness of the scintillator to improve sensitivity (unlike the first configuration), although some broadening of the tracks at the detector is still observed. We will discuss this in more detail in Section~\ref{Configuration-resume}.

This configuration appears to be quite promising as it allows for the maximization of light transfer from the scintillator to the camera while also enabling the camera to be installed outside the experimental chamber, thereby resolving potential issues related to RGA and EMP. Regarding light disturbances from the environment, light tightness can be achieved with this type of structure. However, it is essential to use thin scintillators to avoid deteriorating energy resolution, especially since it will already be degraded due to the use of fiber for which the pitch is larger than those of the camera. If this solution is used, it will be necessary to consider this effect in combination with the significant increase in the number of fibers to transport if one aims to approach the camera’s pitch. For example, the simulation shown in Fig.~\ref{figfibers0.1} and Fig.~\ref{figfibers1} with a 200 µm pitch corresponds to a total of 6,250 fibers. Reducing the pitch to 50 µm (parameter used in IPs analysis) for example, would increase this total to 100,000 fibers.

\subsection{\label{Sc-lens-camera} Scintillator imaged by an optical system on camera\cite{Exp_CSI}}

\begin{figure}[h]
\includegraphics[width=0.5\textwidth]{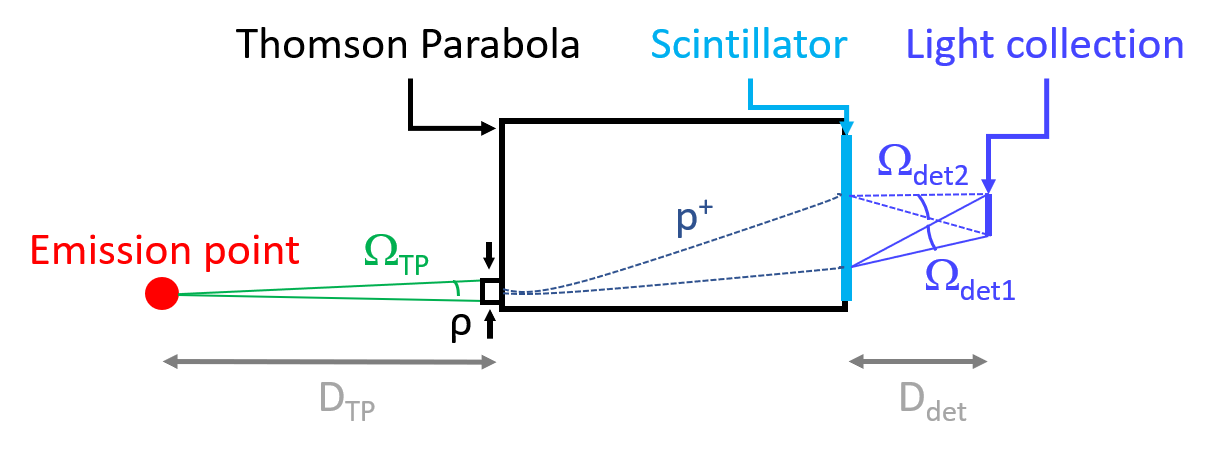}
\caption{\label{fig21} Schematic view of an experiment presenting the various dimensions of interest.}
\end{figure}

This last configuration relies on an optical lens-based system to collect photons emitted from the scintillator and image them on the camera plane (Fig.~\ref{fig21}) with a given detection solid angle according to the localisation of energy deposition. The distance between emission point and TP's pinhole is still the same (30 cm) and we will incorporate a virtual biconvex lens (f = 300 mm, NA = 0.13) at a distance of 30 cm from the scintillator exit plane. Thus, detecting the photons on the lens while recording information about their momentum will allow us to reconstruct, similar to the optical function of the lens, the point at which the photons exit the scintillator. The diameter and distance concerning the lens are arbitrary but correspond to use cases that could be implemented with common lenses and objectives.

\begin{figure}[h]
\includegraphics[width=0.51\textwidth]{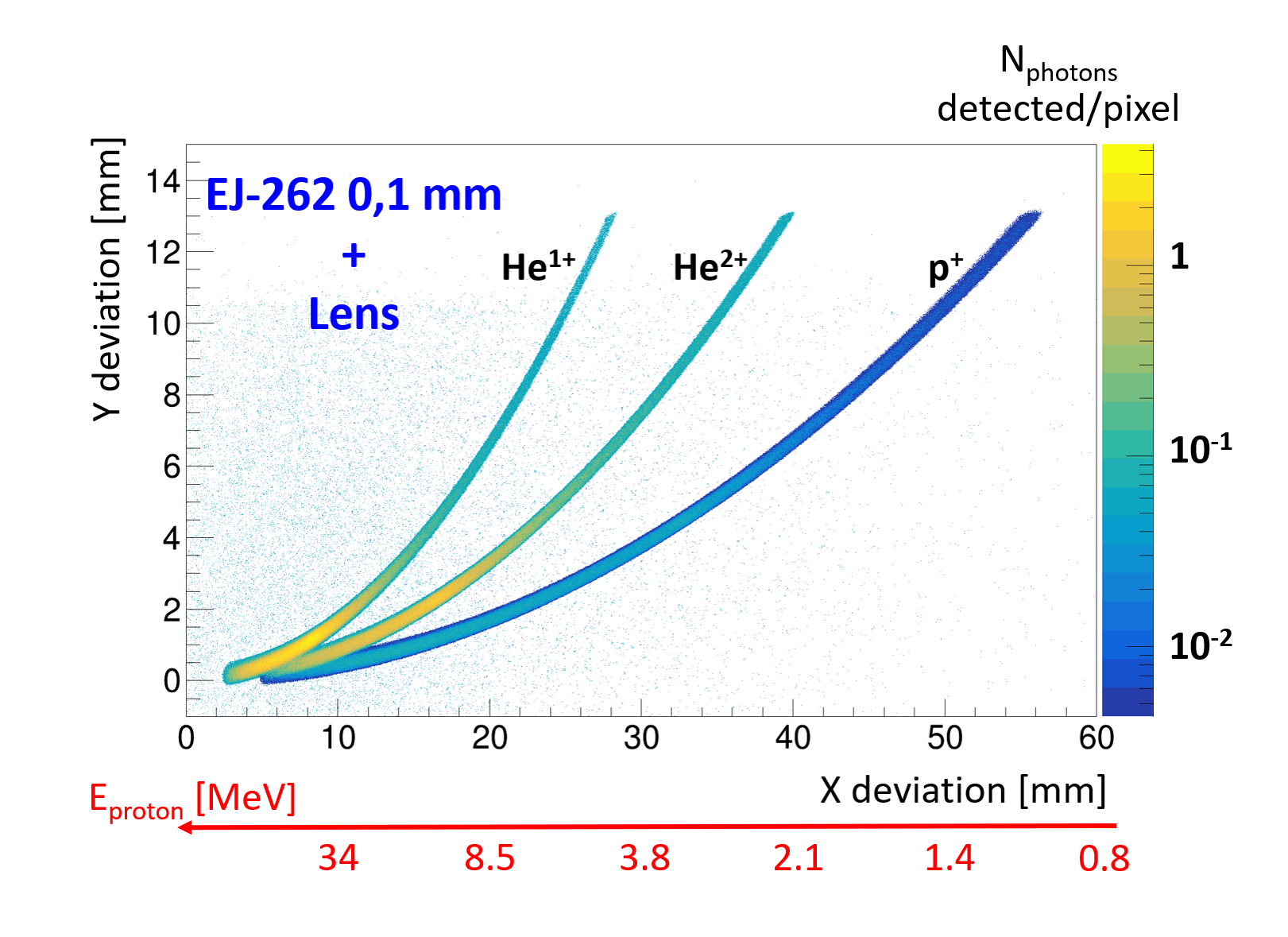}
\caption{\label{figlens0.1}Optical simulations of p$^{+}$, He$^{1+}$ and He$^{2+}$ in the configuration where the photons emitted by a scintillator of 0.1 mm are collected with an optical system. The binning corresponds to the pitch of the camera (6.5 µm).}
\end{figure}

\begin{figure}[h]
\includegraphics[width=0.51\textwidth]{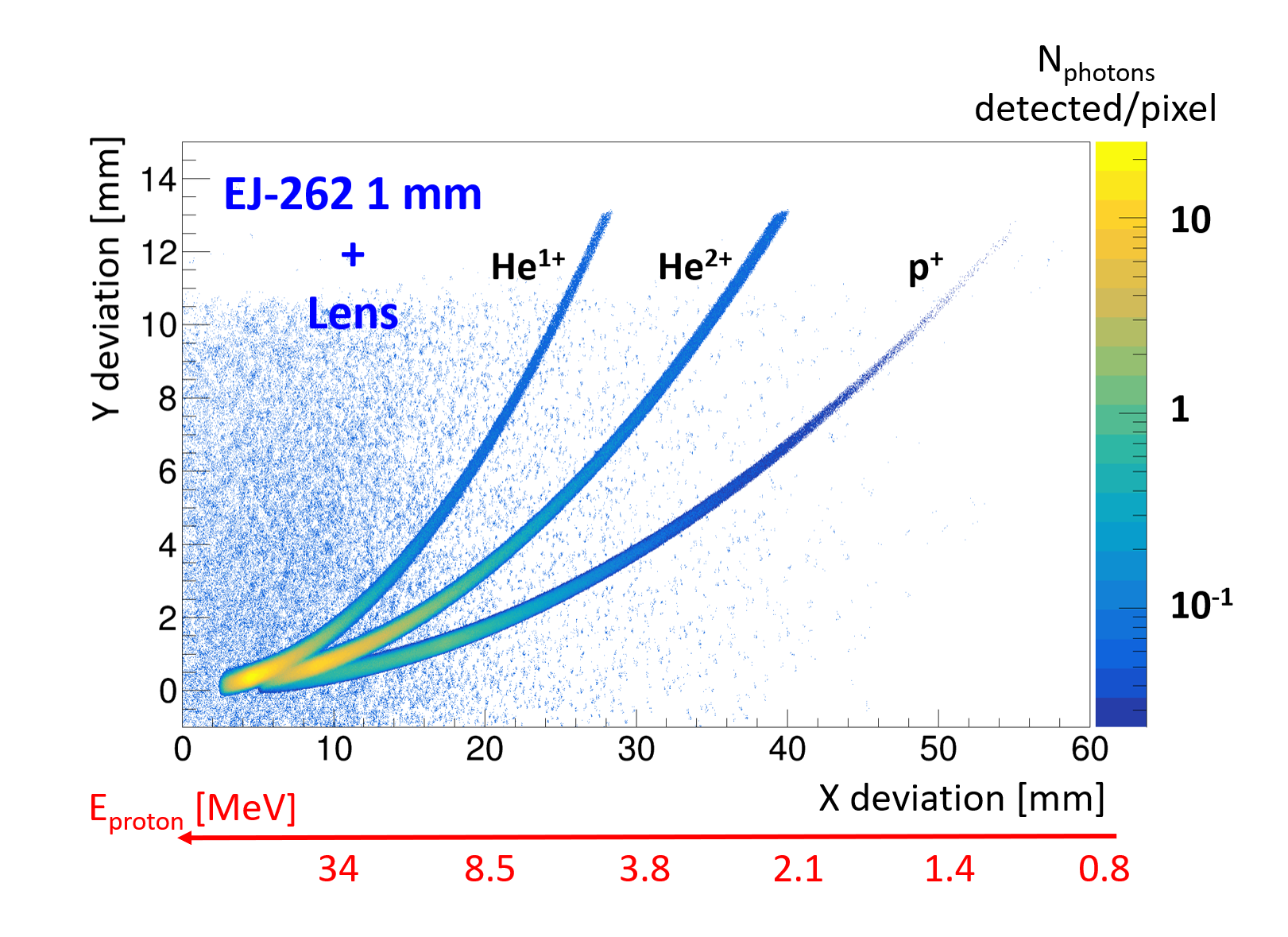}
\caption{\label{figlens1}Optical simulations of p$^{+}$, He$^{1+}$ and He$^{2+}$ in the configuration where the photons emitted by a scintillator of 1 mm are collected with an optical system. The binning corresponds to the pitch of the camera (6.5 µm).}
\end{figure}

The simulation results are shown in Fig.~\ref{figlens0.1} and Fig.~\ref{figlens1} which correspond to scintillators of 0.1 mm and 1 mm respectively.

Unlike the previous configuration, it can be observed that this time, increasing the thickness of the scintillator does not cause any big impact concerning the noise or the broadening of the trace. This can be explained by a simple geometric description of the phenomenon. Indeed, even though photon scattering is more significant for a thicker scintillator, the solid angle of collection of the lens remains the same, which only impacts the number of photons collected by the lens. However, the decrease in the number of photons collected for a thicker scintillator can be compensated for certain energies by the fact that more photons will be produced when ions interact with the scintillator if they do not pass through it completely. We will discuss that in the next section.

In conclusion, This configuration is very interesting because it allows the thickness of the scintillator to be adapted based on the experiment and the expected energy ranges without degrading the energy resolution. However, it is important to keep in mind the loss of light caused at low energies when increasing the thickness of the scintillator. The use of an optical system to displace the light collection is also advantageous in addressing the RGA and EMP issues mentioned in the previous section. Additionally, the displacement would allow the use of filters to eliminate potential light noises by applying spectral filters that transmit light from the scintillator but strongly attenuate other wavelengths. Of course, this kind of filters were not used in our simulations. Nevertheless it is also important to keep in mind that the image obtained on the camera could correspond to a given plane of the scintillator depending on the focal depth of the imaging system. In addition, one must consider the potential significant loss of photons depending on the solid angle of the optical system’s collection, which could be crucial in certain cases. We will return to this point during the tests conducted for the characterization of scintillators in section \ref{Tests_AIFIRA}. 

\subsection{\label{Configuration-resume} Configurations summary}

\begin{figure}[h]
\includegraphics[width=0.48\textwidth]{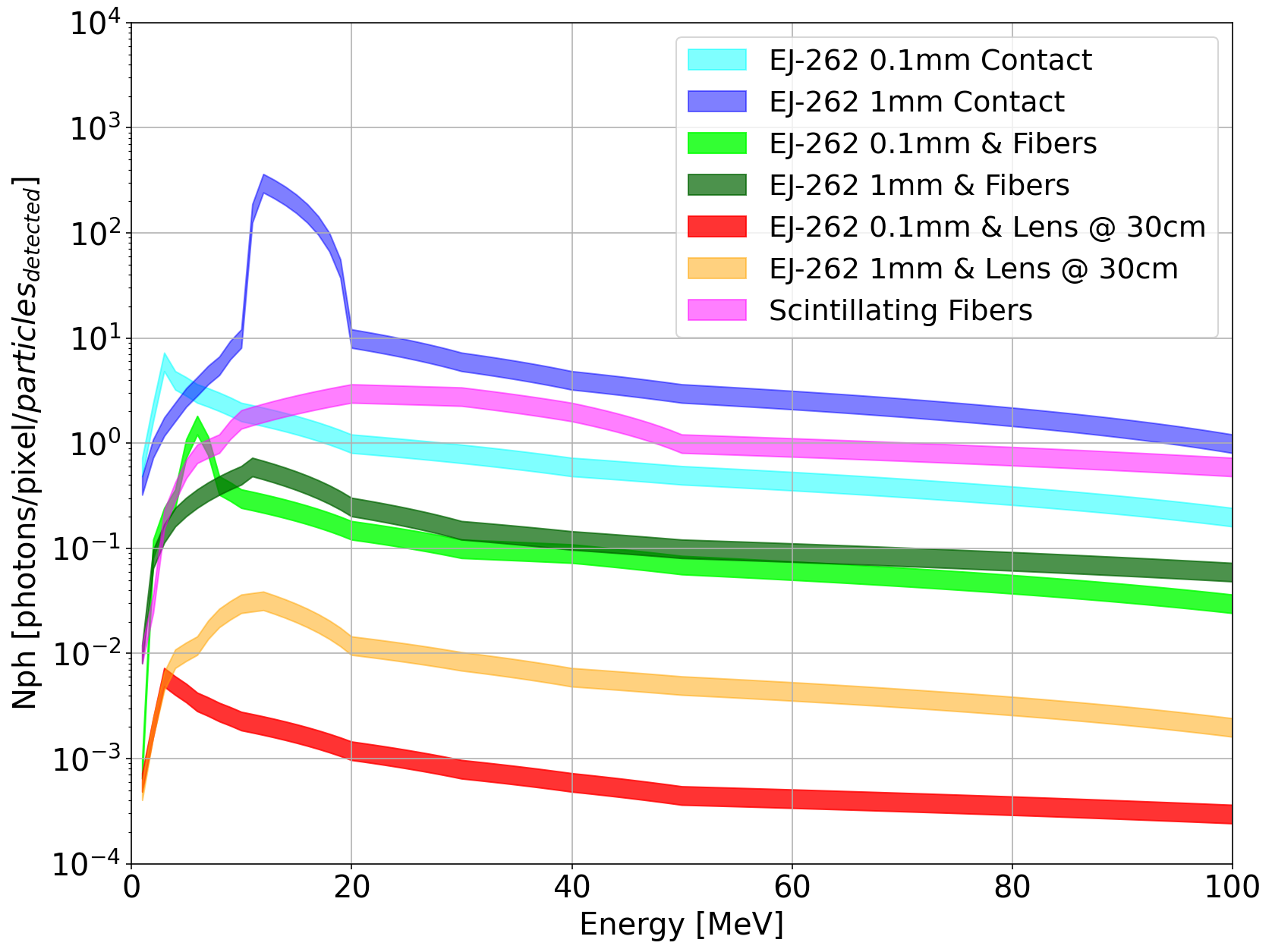}
\caption{\label{fignphglobal}Number of photons collected per pixel and per detected protons as a function of incident proton energy for different thicknesses of EJ-262 scintillators. Width of curves correspond to a uncertainty of 10 $\%$.}
\end{figure}

Now that we have characterized the response of each of our configurations, we will attempt to compare them. To do this, let us begin by comparing the number of photons collected in these different simulations. We obtain Fig.~\ref{fignphglobal}, which shows the evolution of the number of photons detected per pixel and per detected protons as a function of the protons energy for our different simulation configurations. We observe, first, that the amplitude of the range is very large (5 orders of magnitude) between the two extreme configurations, namely the 1 mm scintillator in direct contact with the camera and the light collection from a scintillator via an optical device placed 30 cm away. We also note that, as expected, the configurations with fibers fall within intermediate ranges. It is also worth mentioning the shift in the Bragg peak (around 15 MeV) when the scintillator thickness changes, observed for the two different thickness configurations tested.

However, we observed that the number of collected photons is not a sufficient variable to determine the feasibility of a configuration. Indeed, we noted, particularly for configurations with a large number of detected photons in Fig.~\ref{fignphglobal}, that this was accompanied by a possible degradation of our tracks, and therefore a potential degradation in energy resolution as well as a possible loss of information regarding ion discrimination.

\begin{figure}[h]
\includegraphics[width=0.48\textwidth]{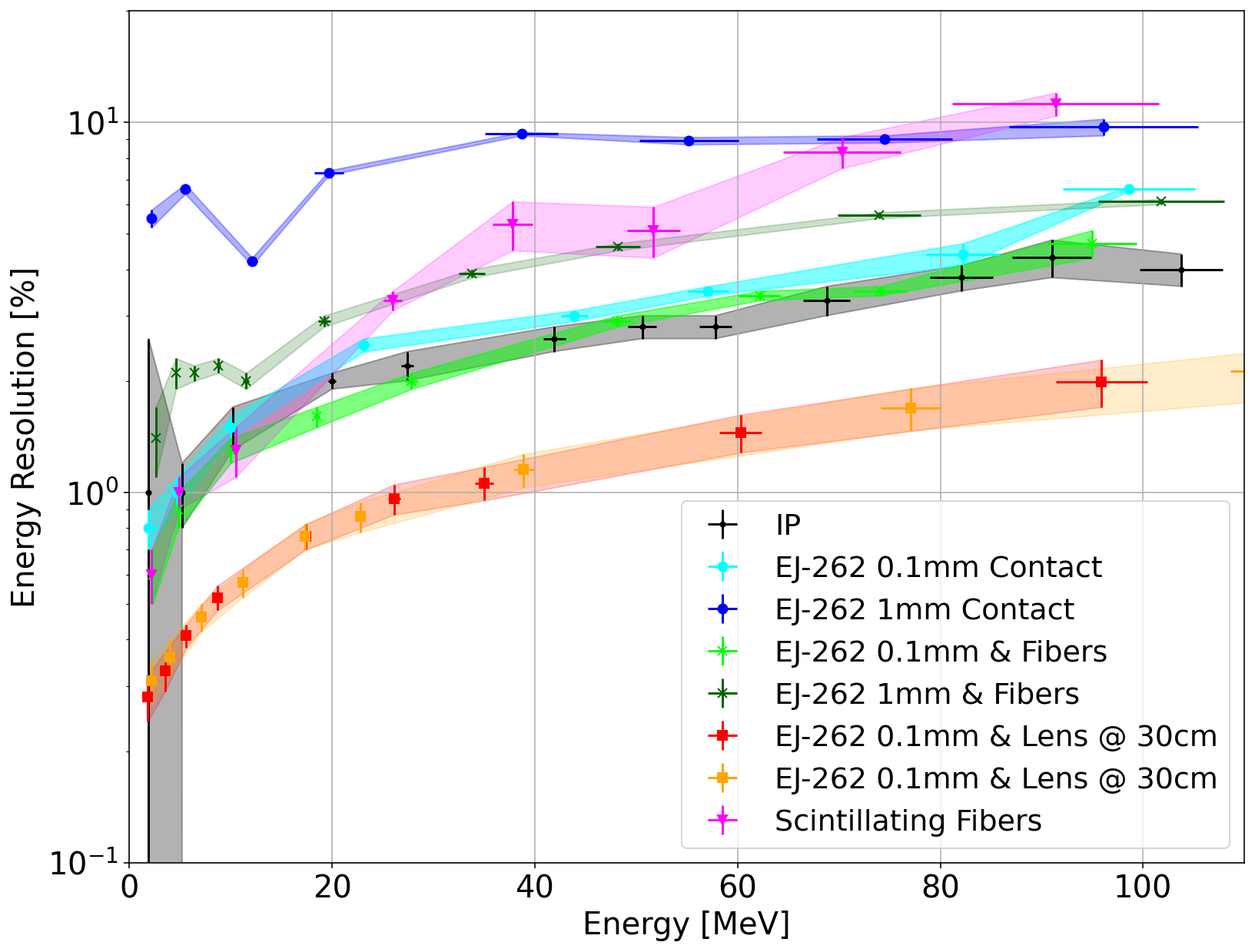}
\caption{\label{figwidthtraces}Energy resolution $\frac{\Delta E}{E}$ expected as a function of incident proton energy for different thicknesses of EJ-262 scintillators and comparison with IP configuration.}
\end{figure}

Thus, Fig.~\ref{figwidthtraces} shows the evolution of the relative energy resolution as a function of the energy of a proton for our different tested configurations. These results are also compared to the most optimistic case achievable with our setup, which corresponds to the configuration with the IPs.

To synthesize the findings from all tested detection configurations, Table~\ref{tab:comparison_configs} summarizes their main characteristics, including signal level, resolution, practical feasibility, and robustness under experimental conditions. This overview facilitates the comparison and highlights trade-offs relevant for experimental design.

\begin{table*}[t]
\centering
\caption{Summary of performance for the tested detection configurations.}
\label{tab:comparison_configs}
\resizebox{\textwidth}{!}{%
\begin{tabular}{|p{4cm}|p{2cm}|p{2.5cm}|p{2.5cm}|p{2.5cm}|p{2.5cm}|}
\hline
\textbf{Configuration} & \textbf{Photon signal} & \textbf{Energy / Spatial resolution} & \textbf{Ease of implementation} & \textbf{EMP / RGA compatibility} & \textbf{Reproducibility} \\
\hline
Scintillator in direct contact with camera & Very high & Degraded (thicker scintillators) & Very easy & Low (camera exposed) & Good, risk of damage \\
\hline
Scintillating fibers imaged by camera & Medium to high & Good (up to 30 MeV), degraded above & Complex (fiber alignment) & Good (camera protected) & Moderate (alignment sensitive) \\
\hline
Light collected via non-scintillating fiber array & Medium & Moderate to good (fiber pitch limited) & Moderately complex & Good & Good \\
\hline
Scintillator imaged through optical lens system & Low to medium & Very good (trace preserved) & Simple to moderate & Very good (camera protected, minimal in-vacuum components) & Excellent \\
\hline
\end{tabular}%
}
\end{table*}

Based on this Table, we can affirm that using a scintillator in direct contact with a camera is hardly feasible (optimistic case without considering proton interactions in the camera, potential damage, and significant background noise). 

We can also observe that it remains preferable, in order to avoid affecting energy resolution, to use scintillators with thicknesses on the order of 100 µm. The gain in detected photon sensitivity is not substantial and could potentially lead to a higher degradation in the scintillators bulk due to the Bragg peak. Favoring thinner scintillators allows us to minimize this effect while maximizing the number of ions passing through.

There are then three configurations, each with its advantages and drawbacks depending on what we aim to measure. Indeed, aside from potential effects at very high energies, fiber scintillators remain an interesting configuration due to their high sensitivity without degrading energy resolution within the considered range. However, the implementation of such a detection system remains the biggest challenge of this configuration due to the management of a large number of fibers. This also applies to the configuration with non-scintillating fibers combined with a scintillator, where the results still fall within the intermediate range. 

The final configuration with an optical lens-based system, despite suffering from potential sensitivity issues, remains the most attractive solution overall, given its ease of implementation and excellent results regarding track quality. This is why, in the section concerning our experimental tests (Section~\ref{Tests_AIFIRA}), we chose to use this configuration.

\subsection{\label{Sc-comparison} Comparative study of scintillator materials}

Now that we have seen the impact of different configurations on scintillation light detection by our system, we will examine, in the context of the configuration with the optical system, the differences observed for the various scintillators mentioned in Table.~\ref{tab:1} (organic, phosphor and inorganic).  It is important to notice that EJ-444 consists of a thin piece of EJ-212 plastic scintillator (0.1 or 1 mm in our case) with a layer ($\approx$ 40 $\mu$m) of silver activated zinc sulfide phosphor (ZnS:Ag) applied to one side. Since we have already characterized the impact of thicknesses for each configuration, we will focus this time on the situation where the scintillator has a thickness of 0.1 mm.

\begin{table*}[t]
    \centering
    \begin{tabular}{|c|c|c|c|c|c|c|}
    \hline
     & EJ-262\cite{EJ262} & EJ-444\cite{EJ444} & YAG:Ce\cite{YAG} \\
    \hline
    Type & Organic & Phosphor (+ Organic) & Inorganic \\
    \hline
    Density & 1.023 & 4.09 (1.023) & 4.57 \\
    $[$g/cm$^3]$ & & & \\
    \hline
    Lightyield & 8 700 & 46 500 (10 000) & 30 000 \\
    $[ph./MeV_{deposited}]$ & & & \\
    \hline
    Max. emission & 481 & 450 (423) & 547 \\
    $[$nm$]$ & & & \\
    \hline
    Sc. attenuation length & 295 & 0.015 (250) & 1000 \\
    @ max. emission $[$cm$]$ & & & \\
    \hline
    Decay time & 2.1 & 200 (2.4) & 70 \\
    $[$ns$]$ & & & \\
    \hline
    Refractive index & 1.58 & 2.36 (1.58) & 1.82 \\
    @ max. emission & & & \\
    \hline
    \end{tabular}
    \caption{Scintillation properties of three different scintillators (organic, phosphorous and inorganic). For EJ-444, values in parentheses correspond to the organic part (EJ-212).}
    \label{tab:1}
\end{table*}

Logically, it is essential to account for the quenching effect of the luminescent sites within a scintillator in order to quantitatively describe the light yield. This phenomenon can be modeled in GEANT4 simulations by incorporating a Birks' constant. The impact of this effect becomes even more significant for heavier particles, as they deposit more energy per unit volume. However, defining an appropriate Birks' constant in GEANT4 is highly dependent on specific simulation parameters and typically requires calibration with experimental data\cite{huber}. For this reason, in the context of this qualitative study, we adopted an ultra-optimistic scenario in which scintillation quenching is neglected.

\begin{figure}[h]
\includegraphics[width=0.5\textwidth]{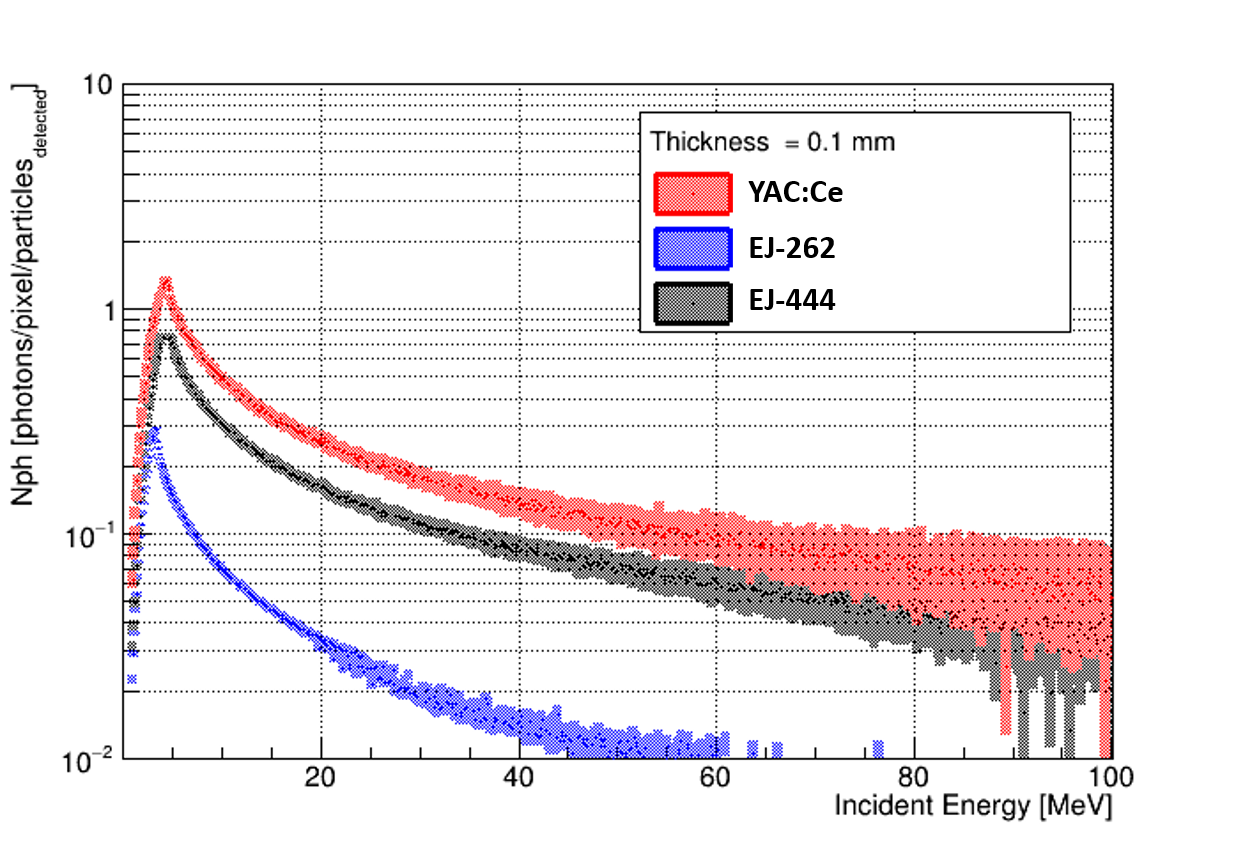}
\caption{\label{fig14}Number of photons collected per pixel and per detected protons as a function of incident proton energy for different scintillators (EJ-262, EJ-444 and YAG:Ce) for a thickness of 0.1 mm in the lens configuration.}
\end{figure}

Thus, we obtain Fig.~\ref{fig14}, which shows the number of photons detected per pixel and per particle that has deposited energy in the scintillator, as a function of the incident energy of the protons for three scintillators (EJ-262, EJ-444, and YAG). It can be easily seen that the maximum energy deposited in the scintillator (which corresponds to the limit case of the proton passing through) is around 4-5 MeV for protons when the scintillator has a thickness of 0.1 mm. As expected, given the light yields of these scintillators, the higher the light yield, the greater the number of photons detected, which favors YAG.

Regarding energy resolution, the simulation does not show significant variations depending on the choice of scintillator. Thus, the energy resolution will mainly be impacted by the configuration and/or the thickness of the scintillator than by the type of scintillator used (Fig.~\ref{figwidthtraces}).

Another point to consider is the response of these scintillators to X/$\gamma$ radiation emitted during the laser shot. Indeed, if these radiations are detected in too large number, they may increase background on the camera, making ion detection more difficult. To provide an answer to this question, we simulated an X/$\gamma$ flash at 30 cm from the detector system following a Boltzmann distribution with a temperature E$_0$ = 805 keV\cite{norreys_gamma_spectrum}. Table.~\ref{tab:2} presents the detection efficiencies obtained for our three scintillators. Naturally, the higher the density and atomic number (Z) of the material, the greater the probability that X/$\gamma$ radiation will interact with the scintillator. This explains why the organic scintillator, made of hydrogen and carbon with a density close to 1, is the most transparent to this radiation, which could prove crucial in certain experiments, despite having a lower yield than YAG:Ce.

Given this point, one may wonder whether it will always be possible, during data analysis, to determine the zero-deflection point using the interaction of X/$\gamma$ rays generated during the laser shot for a plastic scintillator. Simulations indicate that there is a factor of 10 million between the energy deposited by a traversing proton (without applying quenching) and that deposited by an X/$\gamma$ beam following the Boltzmann distribution described earlier. If we are able to detect on the order of 10–100 protons per detection zone, this implies that a minimum gamma flux of 10$^9$ photons on the scintillator would be required to produce a visible image on the detector. However, if such a signal were not observable, the modularity of our TP, combined with the laser alignment system, would still allow us to constrain the origin point. Indeed, the ability to switch detectors without altering the rest of the TP makes it possible to perform a prior calibration using an IP of identical dimensions to locate and define the zero-deflection point.

\begin{table}[h]
    \centering
    \begin{tabular}{|c|c|c|}
    \hline
    Scintillator [0.1 mm] & Density [g/cm$^3$] & $\epsilon_{detection}$ \\
    \hline
    EJ-262 & 1.023 & 0.15 $\%$ \\
    \hline
    EJ-444 & 4.09 (1.023) & 2.55 $\%$ \\
    \hline
    YAG:Ce & 4.57 & 2.64 $\%$ \\
    \hline
    \end{tabular}
    \caption{Estimation of the X/$\gamma$ radiation detection probability for 3 different scintillator (EJ-262, EJ-444 and YAG:Ce). The X/$\gamma$ distribution simulated corresponds to a Boltzmann distribution with E$_0$~=~805~keV.}
    \label{tab:2}
\end{table}

Additionally, we have already mentioned several times the need to pass the RGA tests required to access certain facilities. These tests are due to be conducted soon to determine whether or not these scintillators can be introduced into a vacuum chamber according to the criteria of these facilities even if we already know that it is possible with YAG:Ce scintillator\cite{YAG_Apollon}.

Another important parameter is the cost of these scintillators. While the prices of ELJEN scintillators (EJ-262 and EJ-444) are in the range of several hundred euros depending on size and thickness, the price of a YAG scintillator of the same dimensions is multiplied by a factor of 8-10, which is significant, especially as we currently lack information on the durability of these scintillators when exposed to high repetition rate particle fluxes (other than electrons) in laser-plasma experiments. Appendix \ref{Tests_AIFIRA} will attempt to address this question through tests conducted on the AIFIRA platform of LP2IB to characterize these three scintillators.

\section{\label{Ccl}Conclusion}
In this work, we described the design of a modulable and active Thomson Parabola where all the components have been discussed in detail. The design of the spectrometer is conceived in such a way that the TP elements can be easily adapted to different experimental conditions that may be encountered. In addition, this TP is also built for an easy and quick mounting and includes a built-in alignment device for faster setup and reliable operation in the experiments.

Moreover, through optical simulations conducted using GEANT4, we tested various configurations of "active" detection, and these results allowed us to determine that certain solutions were not feasible. This was particularly true for direct scintillation detection with a contact detector or the use of a bundle of scintillating fibers. 

However, it remains necessary to test our TP in its active mode under real conditions while continuing to develop optical simulations. This will ensure the most comprehensive monitoring of scintillation and guarantee both qualitative and quantitative measurements.

\appendix

\section{\label{Tests_AIFIRA} Scintillators characterization at AIFIRA platform}

\subsection{\label{} Experimental setup}
AIFIRA\cite{AIFIRA} (Applications Interdisciplinaires des Faisceaux d’Ions en Région Aquitaine) is a compact ion beam facility equipped with a single-stage electrostatic accelerator that delivers beams of light ions (protons, deuterons, and helium) in the MeV energy range. Commissioned in 2006, the accelerator (3.5 MV Singletron™ from HVEE, the Netherlands) provides beams with high brightness (20 A m$^{-2}$ rad$^{-2}$ eV$^{-1}$) and excellent energy stability ($\Delta E/E = 10^{-5}$) across five beamlines dedicated to specific applications.

For our tests, we utilized the microbeam line at AIFIRA. This line features a two-stage magnetic quadrupole quintuplet focusing system, and the analysis chamber is equipped with surface-barrier detectors for conducting RBS (Rutherford Backscattering Spectroscopy) and NRA (Nuclear Reaction Analysis) experiments, as well as on-axis Scanning Transmission Ion Microscopy (STIM) for thin targets. The samples are mounted on an XYZ motorized stage with micrometer positioning accuracy. A set of microscope objectives paired with video cameras allows for in situ visualization of the sample at various magnifications. The microbeam can achieve a spot diameter of 1 $\mu$m in high-current mode (several hundred pA) and 0.3 $\mu$m in low-current mode (a few thousand counts per second). A custom-made electrostatic system is employed to raster scan the beam spot across the sample surface for imaging purposes.

For our study, we selected a proton energy of 2.5 MeV to maximize the energy deposition in our 0.1 mm plastic scintillators\cite{Pstar}. The beam size was 1 $\mu$m diameter, and we used fluxes ranging from 0.3 fA to 5 pA. The integration time of the HAMAMATSU camera (Orca-Flash 4.0 LT Plus) was also adjusted (from 3.5 ms to 10 s) to vary the number of integrated protons. Five scintillators were tested:

\begin{itemize}
    \item EJ-262: 0.1 mm and 1 mm.
    \item EJ-444: 0.1 mm and 1 mm (approximately 50 µm ZnS + 0.1/1 mm EJ-212).
    \item YAG:Ce: 0.1 mm.
\end{itemize}

\begin{figure}[h]
\centering
\includegraphics[width=0.41\textwidth]{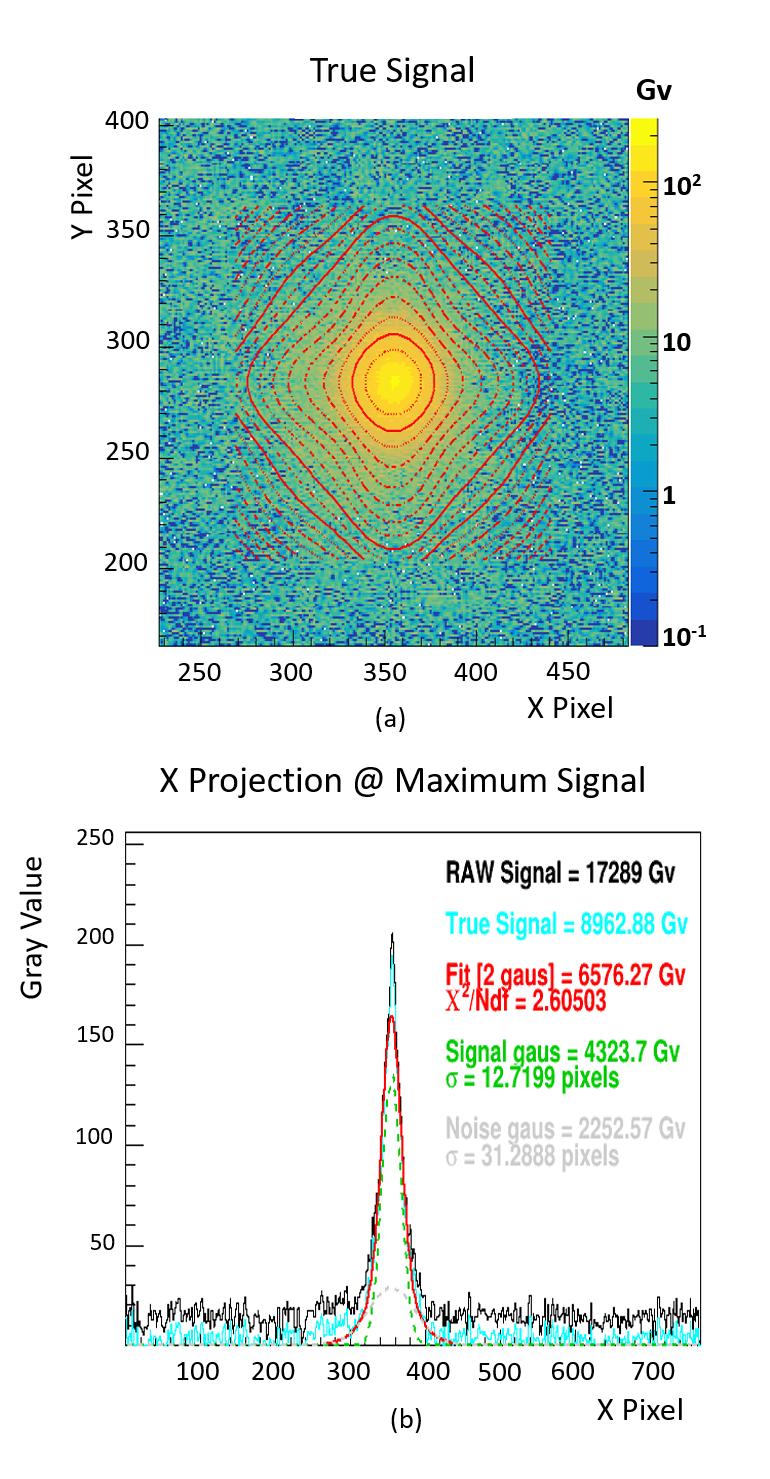} 
\caption{\label{fig17}Example of results obtained with (a) the adjustment of the distribution obtained following the irradiation of a scintillator and (b) the details of the adjustment and the different components when projected onto the x axis at the maximum level signal.}
\end{figure}

The images were recorded in .tiff format and then converted to .csv format (in Gray Value) for analysis using the ROOT software Fig~\ref{fig17}.

A background noise measurement was taken at the beginning of each day to subtract the camera noise from our raw (RAW) data. This allows us to obtain the signal associated solely with scintillation (TRUE). The resulting distribution is then fitted with a double Gaussian in 2D, as shown in Fig~\ref{fig17} (a). Fig~\ref{fig17} (b), which represents the pixel projection along the X-axis of the 2D distribution at the point where the signal is maximum, allows us to observe the different contributions mentioned (RAW signal in black, TRUE signal in cyan, double Gaussian fit in red, Gaussian fit corresponding to the signal in green, and Gaussian fit of the noise caused by light scattering in the scintillator in gray). The fitting enables us to estimate the number of integrated Gray values (Gv) recorded by the camera that correspond to our signal. We will discuss the various observed results in the following section.

\subsection{\label{} Results}
\subsubsection{Measurements behavior for large numbers of protons detected}\mbox{}\\

\begin{figure}[h]
\includegraphics[width=0.48\textwidth]{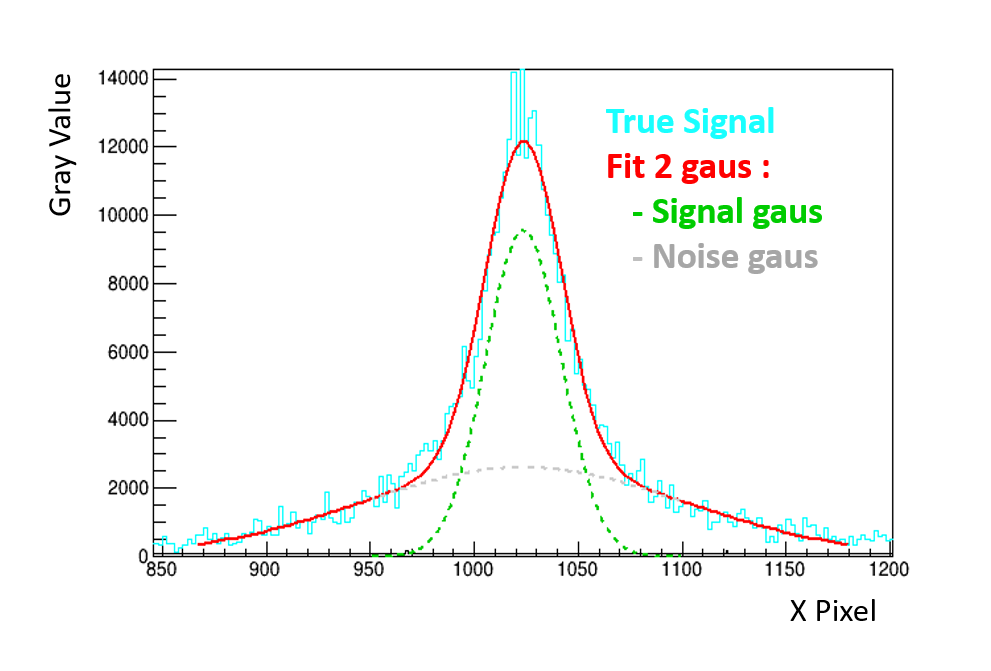}
\caption{\label{fig18} Effects on signal width distribution when a large number of protons are detected.}
\end{figure}

We previously mentioned the need to perform a 2D double Gaussian fit to account for the 'noise' caused by photon scattering within the scintillator. Fig~\ref{fig18}, which corresponds to a higher proton fluxes, will help us visualize this phenomenon by comparing it with Fig~\ref{fig17} (b). Indeed, in the first figure, the maximum number of integrated Gv is around 12,000, compared to 200 in the first example. This means that a much larger number of photons was detected in response to a higher number of protons integrated over the acquisition period. The double Gaussian fit is still well-executed and is even indispensable in this specific case. Since a significantly higher number of photons was produced, more and more detected photons correspond to those that could have scattered within the scintillator, resulting in increased noise caused by these photons, as seen in the gray contribution.

\subsubsection{Scintillation degradation}\mbox{}\\

\begin{figure}[h]
\includegraphics[width=0.49\textwidth]{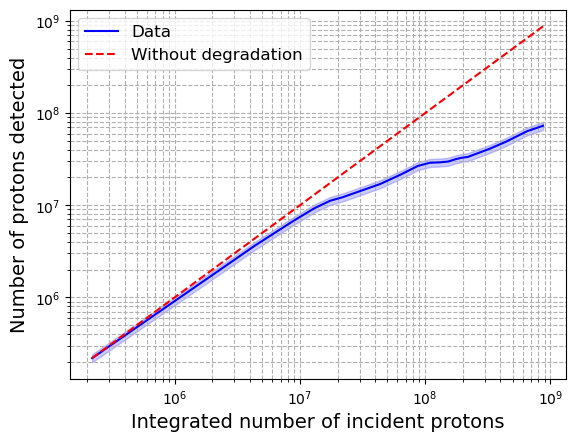}
\caption{\label{fig19} Evolution of the number of protons detected according to the integrated number of incident protons (at a flux of 380 fA) which characterize the degradation of scintillator EJ-262 under proton irradiation on 1 $\mu$m diameter spot.}
\end{figure}

We also took advantage of the proton beam to characterize the degradation of scintillation when the scintillator is exposed to a continuous proton flux of 2.5 MeV at a flux of 380~fA ($\approx$~2.4.10$^6$~p$^+$/s). A long-duration acquisition of 400 seconds was carried out with the HAMAMATSU camera. Regular analysis at fixed intervals of the obtained image enabled us to reconstruct the evolution of the number of integrated Gv as a function of incident protons on the scintillator. The loss of light due to the degradation of the scintillator  will correspond to a lowest number of protons detected than we expect as shown in Fig~\ref{fig19}. For our conditions, the degradation begins to appear around 10$^6$ integrated protons  with a 1 mm beam diameter which is far beyond what the scintillator should be confronted under experimental conditions. The same overall effect was observed with EJ-444, and it will be necessary in a future campaign to also perform this measurement with YAG:Ce.

When performing a test with a flux 1000 times higher on the EJ-262, we even observed permanent degradation of the scintillator with a darkened area in the center of the window. It seems that the scintillator was significantly damaged, particularly due to the high concentration of energy deposition in one spot induced by the Bragg peak. Several explanations for this phenomenon are possible, such as a reduction in optical transmission (due to effects on the plastic based material) and/or a decrease in emission due to the degradation of the scintillating component\cite{Green2011}. The data from this article are in agreement with what we observed with our EJ-262 scintillator.

These findings highlight an important factor when using this type of system in experiments that may be subjected to high particle fluxes, especially for ions. Indeed, as seen in Fig~\ref{fig19}, it is essential and mandatory to regularly calibrate the detector's energy response to correctly estimate the number of incident particles.

\subsubsection{Light response calibration}\mbox{}\\

To eliminate undesirable effects related to variations in scintillation, each measurement was conducted with a brief exposure of around thirty seconds for the scintillator to limit the strong variations at the beginning of irradiation.

\begin{figure}[h]
\includegraphics[width=0.51\textwidth]{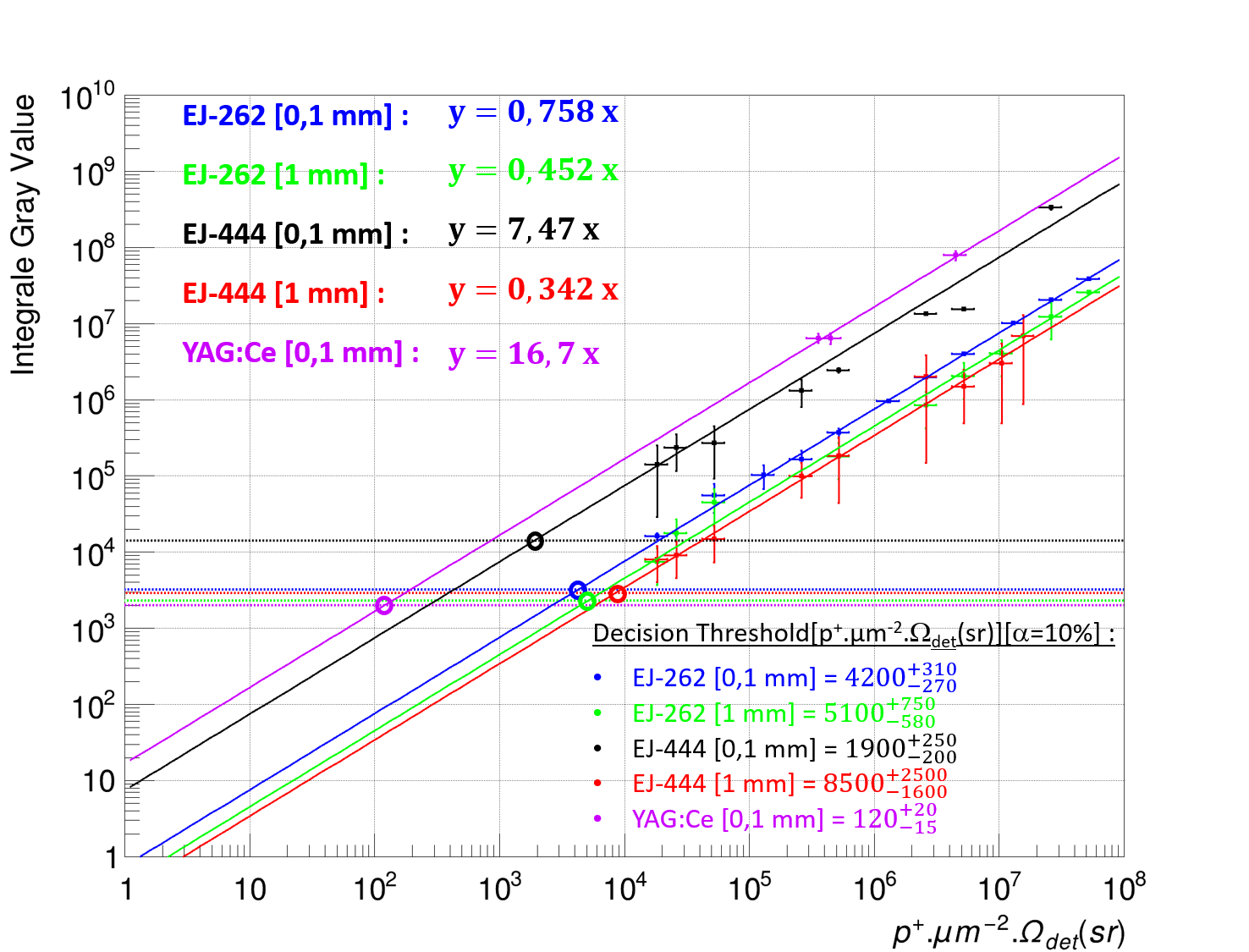}
\caption{\label{fig20} Evolution of the integrated Gray value detected according to the proton flux for our 5 scintillators tested. Limits corresponding to a Decision Threshold with an $\alpha$ value of 10 $\%$ are also displayed.}
\end{figure}

The measurements for our five scintillators are shown in Fig~\ref{fig20}, which represents the number of integrated Gv as a function of the number of protons emitted per unit area and within a solid angle corresponding to that between the scintillator and the lens that collects the light and transmit to the camera. In our case, the number of protons is determined from the flux, the emission area is based on the beam size (1 mm diameter), and the solid angle of detection was approximately 1.88 sr (microscope objective LUCPLFLN 20X used).

The first observation is that all the scintillators' responses are indeed linear. The second observation, as mentioned earlier (see Fig.~\ref{fig14}), clearly shows that YAG is the most efficient scintillator, followed by ZnS (EJ-444), and finally EJ-262. As expected, if the thickness of the scintillator increases, this results in a decrease in the number of detected photons in our specific case where the proton is supposed to deposit all its energy within the first 100 micrometers.

Furthermore, we estimated, based on the dimensions of the spots observed for each scintillator, an upper limit on the number of integrated Gv required to reject the hypothesis that we are only measuring background noise with a confidence level of 90$\%$ \cite{Weise2006BayesianDT}. These limits correspond to the horizontal dashed lines, which then enable us to estimate the minimum flux required to observe a signal when integrating the entire image.

\begin{center}\textbf{Acknowledgement}\end{center}
This study has received financial support from the French State in the framework of
the Investments for the Future programme IdEx université de Bordeaux / GPR LIGHT.

This project has received funding from the European Union's Horizon 2020 research and innovation programme under grant  agreement no. 871124 Laserlab-Europe.

The AIFIRA facility is financially supported by the CNRS, the university of Bordeaux and the Région Nouvelle Aquitaine. We thank the technical staff members of the AIFIRA facility P. Alfaurt and S. Sorieul.

%


\end{document}